Schweizerische Eidgenossenschaft
Confédération suisse
Confederazione Svizzera
Confederaziun svizra

Federal Department of the Environment,
Transport, Energy and Communications DETEC

**Swiss Federal Office of Energy**
Energy Research and Cleantech

Final report from 26 September 2025

# TODC

# Temperature Optimisation in Data Centres

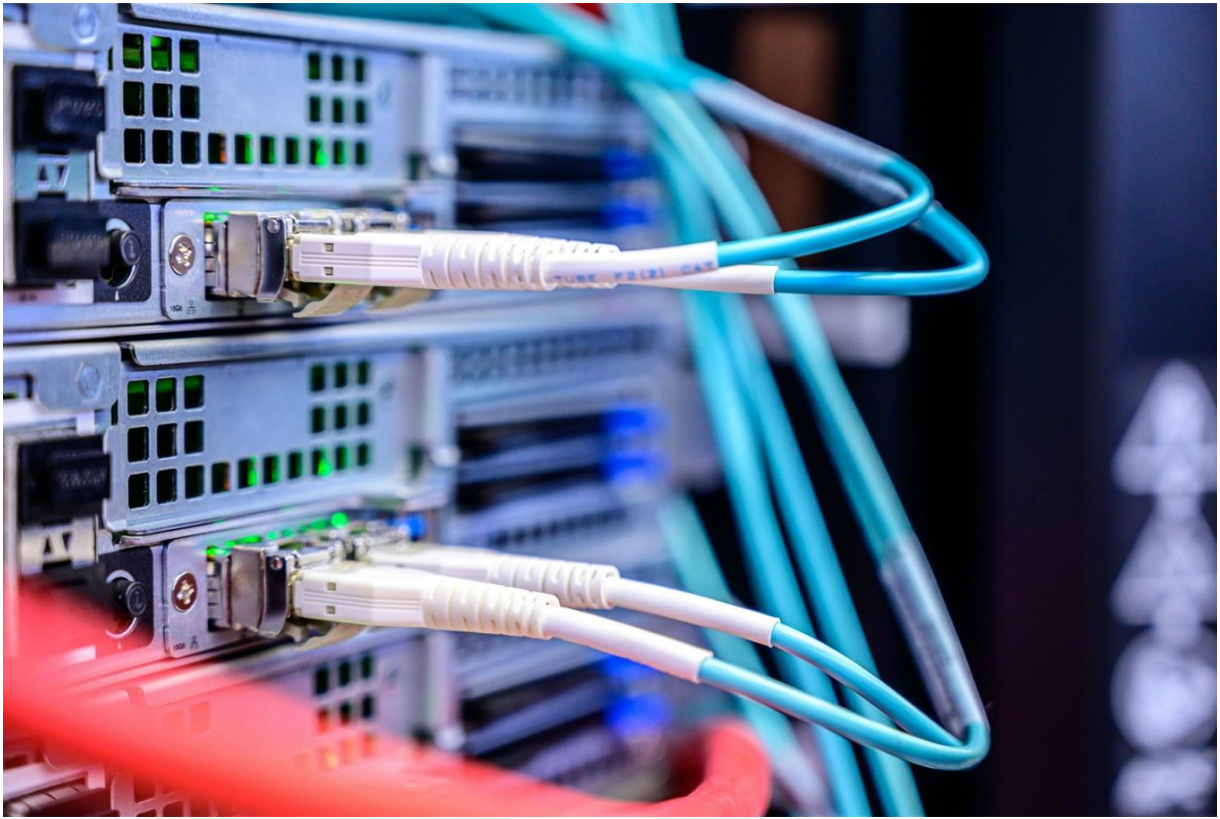

Picture source: Brett Sayles, 2025

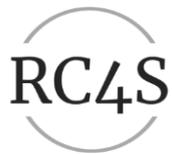

**The author bears the entire responsibility for the content of this report and for the conclusions drawn therefrom.**

# Summary

One factor affecting the efficiency of data centres (DCs) are the server inlet temperatures. Raising these temperatures can significantly enhance the efficiency of cooling systems, by enabling greater use of "free cooling", which uses ambient environmental heat sinks instead of energy-intensive mechanical refrigeration. Higher inlet temperatures, however, can also negatively impact the energy consumption of the IT equipment, primarily by causing internal server fans to work harder.

To which extent these two opposing effects manifest, however, is subject to debate. Recommendations reaching back a decade or more put forward relatively large ranges of recommended temperatures, such as the 18 – 27 °C by the American Society of Heating, Refrigerating and Air-Conditioning Engineers (ASHRAE). Current industry practices are in the upper half of this range, often around 24 – 25 degrees. Theoretical and empiric arguments have been made both in favour of further increasing inlet temperatures – even above 27 degrees – and against it.

Given the conflicting consequences of temperature changes as well as too vague and possibly outdated recommendations, this study set out to analyse how server room consumption depends on inlet temperatures, whether further raising the temperatures brings additional overall benefits or not, and whether widely deployed metrics such as the power usage effectiveness (PUE) are helpful for this assessment.

The analysis was based on academic and industry literature research, a few interviews, but mainly on the case study of a colocation provider in Switzerland. Over the last few years, this case study partner raised server inlet temperatures in two of its data centres. As a result of this partnership, the current study thus had access to detailed primary data on temperatures, server power consumptions, and for one of the two data centres also on the building-wide power consumption over these periods.

The results confirm the positive correlation between inlet temperatures and server power consumption. In this case study, at the higher end of the ASHRAE range and slightly above it, i.e., between 23 – 30 °C, the temperature sensitivity of the server room-wide power consumption was 0.35 – 0.5 percentage per degree centigrade.

Which of the two effects prevails – and, consequently, whether further temperatures increases are meaningful – could not be definitively assessed. Data on the building-wide power consumption was available for one of the data centres. For several temperature changes recorded, however, the results of the analysis were evenly split between significant beneficial and detrimental effects. These conflicting results occurred although several server rooms changed the inlet temperature synchronously, covering substantial shares of the building's overall power consumption. A reasonable expectation had thus been that the building-wide effects would be clear and unambiguous.

The ambiguity might rely on the influence of noise, spurious effects, and confounding factors, which seem more relevant on a building level than they are for single server rooms. A promising way forward could be a controlled experiment, in which temperatures are synchronously modified in all (or at least a substantial share) of a data centre's server rooms, while the computing loads are controlled and kept constant throughout the experiment. This would eliminate what this analysis revealed to be the likely most important confounding factor, i.e., the compute loads in the data centre.

PUE should not be employed in this context for similar analyses. While widely used to assess the energy efficiency of data centres (and especially their cooling infrastructures), PUE is not suited to assess the trade-off between cooling infrastructure and server fans power consumption. For pragmatic reasons, but semantically incorrect, the consumption of server fans is included in the server power consumption, and thus in the wrong part of the PUE fraction. Raising the inlet temperatures will thus always lead to a PUE decrease, both because cooling power consumption will decrease and server fan (and thus seemingly server) power consumption will increase. Such PUE decrease says nothing about which of the two effects is stronger, and might thus easily be misleading.

For now, however, a few indications point towards optimal inlet temperatures at the upper end of the ASHRAE recommendation (around 25–27 °C), which would also corroborate with the evidence from hyperscale data centres, and the opinion of an interviewed expert surveyed.

# Zusammenfassung

Ein Faktor, der die Effizienz von Rechenzentren (RZ) beeinflusst, sind die Server-Eingangstemperaturen. Eine Erhöhung dieser Temperaturen kann die Effizienz der Kühlsysteme erheblich steigern, indem sie eine stärkere Nutzung der „freien Kühlung" ermöglicht, bei der Umgebungswärmesenken anstelle von energieintensiven mechanischen Kälteanlagen genutzt werden können. Höhere Eingangstemperaturen können sich jedoch auch negativ auf den Energieverbrauch der IT-Ausrüstung auswirken, vor allem dadurch, dass die internen Serverlüfter stärker arbeiten müssen.

In welchem Ausmass sich diese beiden gegensätzlichen Effekte manifestieren, ist jedoch umstritten. Empfehlungen, die ein Jahrzehnt oder mehr zurückreichen, schlagen relativ weite Temperaturbereiche vor, wie etwa die 18 – 27 °C der American Society of Heating, Refrigerating and Air-Conditioning Engineers (ASHRAE). Die gängige Branchenpraxis liegt in der oberen Hälfte dieses Bereichs, oft bei 24 – 25 Grad. Über die Zeit wurden sowohl theoretische wie auch empirische Argumente für und gegen eine weitere Erhöhung der Eingangstemperaturen – sogar über 27 Grad hinaus – vorgebracht.

Angesichts der widersprüchlichen Effekte von Temperaturänderungen sowie vager und möglicherweise veralteter Empfehlungen, analysiert diese Studie, wie der Server-Stromverbrauch von den Eingangstemperaturen abhängt, ob eine weitere Erhöhung der Temperaturen Gesamtvorteile bringt oder nicht und ob verbreitete Metriken wie die Power Usage Effectiveness (PUE) in der Bewertung hilfreich sind.

Die Analyse basierte auf akademische und industrielle Fachliteratur, einigen Interviews, aber hauptsächlich auf der Fallstudie eines Colocation-Anbieters in der Schweiz. In den letzten Jahren hat dieser die Server-Eingangstemperaturen in zwei seiner Rechenzentren erhöht. Die vorliegende Studie hatte somit Zugang zu detaillierten Primärdaten zu Temperaturen, Server-Stromverbrauch und, für eines der beiden Rechenzentren, auch zum gebäudeweiten Stromverbrauch während dieses Zeitraums.

Die Ergebnisse bestätigen die positive Korrelation zwischen Eingangstemperaturen und dem Stromverbrauch der Server. In dieser Fallstudie betrug die Temperaturempfindlichkeit des Stromverbrauchs des gesamten Serverraums am oberen Ende des ASHRAE-Bereichs und leicht darüber (d.h. zwischen 23 – 30 °C), 0,35 – 0,5 Prozent Strommehrverbrauch pro Grad Celsius.

Welcher der beiden Effekte überwiegt – und ob folglich weitere Temperaturerhöhungen sinnvoll sind – konnte nicht abschliessend beurteilt werden. Daten zum gebäudeweiten Stromverbrauch waren für eines der RZ verfügbar. Bei mehreren aufgezeichneten Temperaturänderungen waren die Analyseergebnisse jedoch gleichmässig zwischen signifikant vorteilhaften und nachteiligen Effekten aufgeteilt. Diese widersprüchlichen Ergebnisse traten auf, obwohl mehrere Serverräume die Eingangstemperatur synchron änderten und dabei einen wesentlichen Anteil am Gesamtstromverbrauch des Gebäudes ausmachten. Eine vernünftige Erwartung war daher, dass die gebäudeweiten Effekte klar und eindeutig sein würden.

Die Mehrdeutigkeit könnte auf den Einfluss von Rauschen, Scheineffekten und Störfaktoren zurückzuführen sein, die auf Gebäudeebene relevanter zu sein scheinen als für einzelne Serverräume. Ein vielversprechender Lösungsansatz könnte ein kontrolliertes Experiment sein, bei dem die Temperaturen in allen (oder zumindest einem wesentlichen Teil der) Serverräume eines Rechenzentrums synchron verändert werden, während die Rechenlasten kontrolliert und während des gesamten Experiments konstant gehalten werden. Dies würde den Faktor eliminieren, der sich in dieser Analyse als der wahrscheinlich wichtigste Störfaktor herausgestellt hat, nämlich die Rechenlasten im Rechenzentrum.

PUE sollte in diesem Zusammenhang nicht zur Analyse herangezogen werden. Obwohl PUE zur Bewertung der Energieeffizienz von RZ (und insbesondere ihrer Kühlinfrastrukturen) weithin verwendet wird, ist die Metrik ungeeignet, den Kompromiss zwischen Verbrauch der Kühlinfrastruktur und der Serverlüfter zu bewerten. Aus pragmatischen, aber semantisch falschen Gründen wird der Verbrauch der Serverlüfter in den Stromverbrauch der Server einbezogen und landet somit im falschen Teil des PUE-Bruchs. Eine Erhöhung der Eingangstemperaturen führt daher immer zu einer PUE-Senkung, da sowohl der Stromverbrauch der Kühlung sinkt als auch der Stromverbrauch der Serverlüfter (und damit scheinbar auch der Server) steigt. Eine solche PUE-Senkung informiert jedoch nicht, welcher Effekt stärker ist, und kann daher leicht irreführend sein.

Solange eine genaue Antwort aussteht, deuten einige Anzeichen darauf hin, dass die optimalen Einlasstemperaturen am oberen Ende der ASHRAE-Empfehlung (etwa 25–27 °C) liegen könnten; dies würde auch mit den Erkenntnissen aus Hyperscale-Rechenzentren übereinstimmen, wie auch mit der Meinung eines befragten Experten.

# Résumé

Un des facteurs influençant l'efficacité des centres de données (CD) est la température d'entrée des serveurs. Augmenter ces températures peut améliorer de manière significative l'efficacité des systèmes de refroidissement, en permettant une plus grande utilisation du « refroidissement naturel » (free cooling), qui exploite les dissipateurs thermiques de l'environnement ambiant au lieu de la réfrigération mécanique énergivore. Cependant, des températures d'entrée plus élevées peuvent également avoir un impact négatif sur la consommation d'énergie de l'équipement informatique, principalement en obligeant les ventilateurs internes des serveurs à travailler plus intensément.

Toutefois, la mesure dans laquelle ces deux effets opposés se manifestent fait l'objet de débats. Des recommandations datant d'une décennie ou plus proposent des plages de températures recommandées relativement larges, comme les 18 – 27 °C préconisés par l'American Society of Heating, Refrigerating and Air-Conditioning Engineers (ASHRAE). Les pratiques actuelles de l'industrie se situent dans la moitié supérieure de cette plage, souvent autour de 24 – 25 degrés. Des arguments théoriques et empiriques ont été avancés tant en faveur qu'en défaveur d'une nouvelle augmentation des températures d'entrée, même au-delà de 27 degrés.

Compte tenu des conséquences contradictoires des changements de température ainsi que des recommandations trop vagues et potentiellement obsolètes, cette étude a entrepris d'analyser la manière dont la consommation d'une salle de serveurs dépend des températures d'entrée, si une nouvelle augmentation des températures apporte ou non des avantages globaux supplémentaires, et si des indicateurs largement répandus comme le Power Usage Effectiveness (PUE) sont utiles pour cette évaluation.

L'analyse s'est basée sur des recherches dans la littérature académique et industrielle, quelques entretiens, mais principalement sur l'étude de cas d'un fournisseur de colocation en Suisse. Au cours des dernières années, ce partenaire de l'étude de cas a augmenté les températures d'entrée des serveurs dans deux de ses centres de données. Grâce à ce partenariat, la présente étude a donc eu accès à des données primaires détaillées sur les températures, la consommation électrique des serveurs et, pour l'un des deux centres de données, également sur la consommation électrique de l'ensemble du bâtiment sur ces périodes.

Les résultats confirment la corrélation positive entre les températures d'entrée et la consommation électrique des serveurs. Dans cette étude de cas, à l'extrémité supérieure de la plage de l'ASHRAE et légèrement au-delà, c'est-à-dire entre 23 et 30 °C, la sensibilité de la consommation électrique de l'ensemble de la salle de serveurs à la température était de 0,35 à 0,5 pour cent par degré centigrade.

Il n'a pas été possible de déterminer lequel des deux effets prévaut – et, par conséquent, si de nouvelles augmentations de température sont judicieuses. Des données sur la consommation électrique de l'ensemble du bâtiment étaient disponibles pour l'un des centres de données. Cependant, pour plusieurs changements de température enregistrés, les résultats de l'analyse étaient répartis de manière égale entre des effets bénéfiques et préjudiciables significatifs.

Ces résultats contradictoires sont apparus bien que plusieurs salles de serveurs aient modifié la température d'entrée de manière synchrone, représentant une part substantielle de la consommation électrique globale du bâtiment. On aurait donc pu raisonnablement s'attendre à ce que les effets à l'échelle du bâtiment soient clairs et sans ambiguïté.

L'ambiguïté pourrait provenir de l'influence du bruit, d'effets parasites et de facteurs de confusion, qui semblent plus pertinents à l'échelle d'un bâtiment que pour des salles de serveurs individuelles. Une voie prometteuse pourrait être une expérience contrôlée, dans laquelle les températures seraient modifiées de manière synchrone dans toutes les salles de serveurs d'un centre de données (ou du moins dans une part substantielle d'entre elles), tandis que les charges de calcul seraient contrôlées et maintenues constantes tout au long de l'expérience. Cela éliminerait ce que cette analyse a révélé comme étant le facteur de confusion probablement le plus important, à savoir les charges de calcul dans le centre de données.

Le PUE ne devrait pas être utilisé dans ce contexte pour des analyses similaires. Bien que largement utilisé pour évaluer l'efficacité énergétique des centres de données (et en particulier de leurs infrastructures de refroidissement), le PUE n'est pas adapté pour évaluer le compromis entre la consommation d'énergie de l'infrastructure de refroidissement et celle des ventilateurs de serveurs. Pour des raisons pragmatiques, mais sémantiquement incorrectes, la consommation des ventilateurs de serveurs est incluse dans la consommation d'énergie des serveurs, et donc dans la mauvaise partie de la fraction du PUE. L'augmentation des températures d'entrée entraînera donc toujours une diminution du PUE, à la fois parce que la consommation d'énergie pour le refroidissement diminuera et que celle des ventilateurs de serveurs (et donc apparemment des serveurs) augmentera. Une telle diminution du PUE ne dit rien sur lequel des deux effets est le plus fort ni sur la consommation d'énergie globale du centre de données, et pourrait donc être facilement trompeuse.

En attendant une réponse précise, certains indices suggèrent que les températures d'entrée optimales pourraient se situer dans la fourchette supérieure recommandée par l'ASHRAE (environ 25 à 27 °C) ; cela corroborerait également les preuves recueillies dans les centres de données hyperscale et l'avis d'un expert interrogé.

# Main findings («Take-Home Messages»)

- Raising server inlet temperatures tends to increase the server power consumption due to heightened server fan activity needed to compensate for the warmer air, and to decrease cooling infrastructure power consumption due to less intense and possibly less frequent chiller activity.

- Relying on primary data from two data centres of a colocation provider in Switzerland, this study succeeds in confirming and quantifying the first effect: In the upper half of the ASHRAE recommended temperature range and slightly above it, i.e., between 23 – 30° C, the server power consumption correlated positively with the temperature, with a server room-wide temperature sensitivity of 0.35 – 0.5 % / °C.

- Based on the available data and due to ambivalent results, the study did not succeed in answering which effect is stronger for the entire data centre: the increase in server consumption or the savings in cooling consumption. Whether raising inlet temperatures beyond the ASHRAE upper limit (i.e., 27 °C) yields additional overall benefits thus remains an open question that would be best addressed through controlled experiments.

- A few pieces of evidence indicate nevertheless that the upper part of the current ASHRAE recommendation (i.e., around 25 – 27 °C) might be the current sweet spot for colocation DCs.

- The widely used power usage effectiveness (PUE) metric is not at all useful in this context. For practical reasons but semantically wrong, the server fan power consumption is included in the consumption of servers (and thus, of the IT infrastructure) and is thus on the wrong side of the PUE fraction. An inlet temperature raise, which will lead to increased fan consumption, will thus always yield a PUE decrease, irrespective of what happens to the overall data centre power consumption.

# Contents

# List of figures

# List of tables

# List of abbreviations

| | |
|---|---|
| AI | artificial intelligence |
| ASHRAE | American Society of Heating, Refrigerating and Air-Conditioning Engineers |
| BTU | British thermal units (a measure of energy) |
| CAGR | compound annual growth rate |
| COP | coefficient of performance |
| CRAC | computer room air conditioner |
| CRAH | computer room air handler |
| DC | Data Centre |
| GPU | graphics processing unit (a type of processor deployed for accelerated computing, initially used in computer graphics and now increasingly in machine learning algorithms) |
| HVAC | Heating, Ventilation, and Air Conditioning |
| ML | machine learning |
| PUE | power usage effectiveness |
| RDHx | rear-door heat exchanger |
| RISE | Research Institute of Sweden |
| ROI | return on investment |
| SFOE | Swiss Federal Office of Energy |
| TDP | thermal design power |
| TIM | thermal interface material |
| TPU | tensor processing unit (a custom-built AI accelerator, developed and deployed by Google/Alphabet for machine learning algorithms) |
| UPS | uninterruptable power supply |

# 1 Introduction

## 1.1 Context and motivation

One of the most influential factors affecting cooling energy consumption in data centres (DCs) is the operating temperature within server rooms, specifically the server inlet temperature. Raising this temperature can significantly enhance the efficiency of cooling systems and enable greater use of "free cooling" or economisation techniques, which use ambient environmental conditions instead of energy-intensive mechanical refrigeration. However, higher inlet temperatures can also impact the energy consumption of the IT equipment itself, primarily by causing internal server fans to work harder. In the worst case, it can lead to overheating, which in turn can result in hardware malfunction, data loss and downtime. On the other hand, excessive cooling of servers can lead to increased energy and water consumption – and therefore increased environmental and operating costs – without significant benefits. It is therefore important to find the right balance for server inlet temperature.

Industry standards and recommendations dating back 10 to 15 years, such as those from Dell (Moss, 2009) and IBM (IBM, 2015), recommend relatively low inlet temperatures. Likewise, the current guidelines and best practice document from the American Society of Heating, Refrigerating and Air-Conditioning Engineers (ASHRAE) also recommends an inlet temperature of 18° – 27° C (ASHRAE TC9.9, 2016); however, this document from 2016 is also almost a decade old.

Based on these and similar documents, the European Commission's 2019 "ecodesign requirements for servers and data storage products" (European Commission, 2019) recommends 27° C as the ideal inlet temperature for operating data centres. The document states that although cooling energy would further decrease as the temperature increases, fan energy would increase disproportionately: "Although increasing the temperature settings leads to a reduction in HVAC consumption, it also has the opposite effect on the servers themselves, where consumption increases as the internal fans blow harder. This results in a 'sweet spot' temperature setting of 27°C, where the sum of HVAC consumption and server internal fans is at its lowest. Operating at temperatures higher than 27°C would therefore lead to an increase in energy consumption in the data centre, as the internal server fans increase their performance" (Petithuguenin, Clinger and Leddy, 2023).

## 1.2 Project objective

This latter study aimed to review and update the European Commission's ecodesign directives, including the recommended temperature. The US Energy Star eco-label also recommends increasing the temperature towards the upper end of the ASHRAE range, although without settling on a concrete number (Energy Star, 2024). And hyperscale operators seem to indeed operate their datacentres around this limit or even above it (Swinhoe, 2024).

The increasing power density of chips and servers, however, might also induce an opposing trend. The latest generation of GPUs require substantially more power than previous generations; Nvidia's latest Blackwell generation of GPUs B200, for example, has a thermal design power' (TDP) of 1.2 kW on its own and 2.7 kW as a GB200 superchip, which combines 2 Blackwell GPUs with a Grace CPU (Smith, 2024). Probably due to this high power density, Nvidia advises an operating range of 5 – 30°C for the B200 chip (Nvidia, 2024), the lower margin of which is substantially lower than typical ASHRAE recommendations.

In this context, the current report aims to provide an analysis of the optimal server inlet air temperature range that minimises the total energy consumption – encompassing both the energy used by the IT equipment (including servers, storage, networking, and their internal fans) and the energy used by the supporting cooling infrastructure (chillers, pumps, air handlers, cooling towers, etc.) – specifically within the context of modern large (i.e., hyperscale and colocation) data centres.

# 2 Background

## 2.1 Increasingly dense heat production in server rooms

The power consumption of both general-purpose as well as dedicated servers has been continuously growing over the past decade. A recent study found that for newly released general-purpose servers, the average maximum power consumption (i.e., the average of the maximum power consumptions of individual servers) grew between 2015 and 2023 as follows:

- for single-CPU servers from 115 W to 307 W,
- for dual-CPU servers from 286 W to 704 W, and
- for servers with more than two CPUs from 799 W to 1,742 W (Coroamă *et al.*, 2025).

The graphics processing units (GPUs) and tensor processing units (TPUs) deployed in machine learning (ML) algorithms saw an even faster development, the average thermal design power (TDP) of newly released GPUs and TPUs increasing in the decade 2013 – 2023 from an average of 191 W to 1,183 W (Coroamă *et al.*, 2025). The corresponding compound annual growth rates (CAGRs) are 14-15% for the general-purpose server categories and 20% for hardware accelerators (GPUs and TPUs), respectively.

The steadily increasing power consumption of both general-purpose servers and accelerated computing chips entails a similar trend towards an increasing power density of servers, and correspondingly server cabinets and server rooms. Consequently, the past years witnessed a continuous increase of the heat that needs to be removed from server rooms.

## 2.2 Heat removal from data centres

This need has led to continuously improving and specialised paradigms for the cooling systems. Some 30 years ago, the heating, ventilation, and air conditioning (HVAC) of data centres resembled that of office buildings (Patel *et al.*, 2025). In modern DCs that generate massively larger amounts of heat per surface, such paradigm would not only be extremely inefficient, but it would simply not be possible to remove the heat with such simplistic systems.

Various cooling innovations have thus emerged over the past decades for data centres. They start from the chip level (how to quickly dissipate the heat from the chip) to the server level, room and ultimately building level. This subsection provides a quick overview of the current state-of-the-art; a detailed description can be found, for example, in (Patel *et al.*, 2025).

### 2.2.1. Chip and server level

Heat removal begins at the smallest scale, where the heat originates. There, the electrical energy consumed by the computing chips and other IT equipment is converted into heat. In traditional air-cooled servers, a thermal interface material (TIM) is placed directly atop the die to transfer the heat from the chip to the outside. This is called a *heat spreader*, its primary job being to take the concentrated heat generated by the very small and powerful chip die and distribute ("spread") it over a larger surface area. This creates a more uniform temperature across the top of the chip package, making it easier for the cooling solution to work effectively. What is often referred to as *heat sink* is mounted on top of the heat spreader. It is typically made of a highly conductive metal like aluminium or copper and features a design with many fins or pins. This large surface area is specifically designed to dissipate the heat that has been spread by the TIM into the surrounding air (Schelling, Shi and Goodson, 2005).

The name "sink" is perhaps not ideal, as this is only a temporary heat storing device and not the final environmental heat sink (which is often the atmosphere or a body of water, as will be discussed in Section 2.2.2). In the context of chip architectures, however, "heat sink" is the established terminology. Spreaders increase the surface of contact between the electronic component and the heat sink, which reduces the heat flux (heat per area and time) and hence improves cooling performance. Additionally, the heat sink conduces the heat away from the active components of the chip (Schelling, Shi and Goodson, 2005).

Server fans are then responsible for removing the heat from the heat sinks of the IT components inside servers. The primary purpose of server fans is to cool components inside the server chassis, especially the chips. Sufficient airflow is required, often around 165 to 170 cubic feet per minute – i.e., 4.7 – 4.8 m$^3$ per minute – per kW of power (Patel *et al.*, 2025). The exact amount, however, depends not only on the server's power but also on the server inlet temperature: A lower inlet temperature makes the airflow more efficient at absorbing heat and thus a lower airflow (and thus server fan speed) is required. Conversely, as the air temperature entering the server increases, server fans typically need to speed up to increase the airflow (Torrell, Brown and Avelar, 2016).

2.2.2. Server room and building level

Figure 1 provides a reference heat removal design for modern data centres on server room and building level. It consists of four cooling circuits, which together remove the heat away from the servers and ultimately to an environmental heat sink. The four circuits comprise i) the in-server-room air circuit, which uses air to remove the heat away from the servers and transfers its heat in the CRAC/CRAH unit to the water of ii) the chilled water circuit, which takes the heat out of the room to the chiller, iii) the chiller refrigerant circuit, which is between the chiller's evaporator and condenser, and finally iv) the condenser water circuit, which takes the heat out of the data centre to the environmental hat sink. The individual components are discussed below Figure 1.

Starting from this generic reference design, there are numerous further variations possible. Their detailed analysis is out of scope for this report, which only briefly mentions some main variations: When free cooling can be used, for example, the chiller is not needed. For liquid cooling, the air circuit is replaced by another liquid circuit, which can be independent or an extension of the chilled water circuit. The final heat sink must not be the atmosphere but can be a body of water. Or it can also be an artificial final heat sink, e.g. when the DC cooling heat is being reused in district heating.

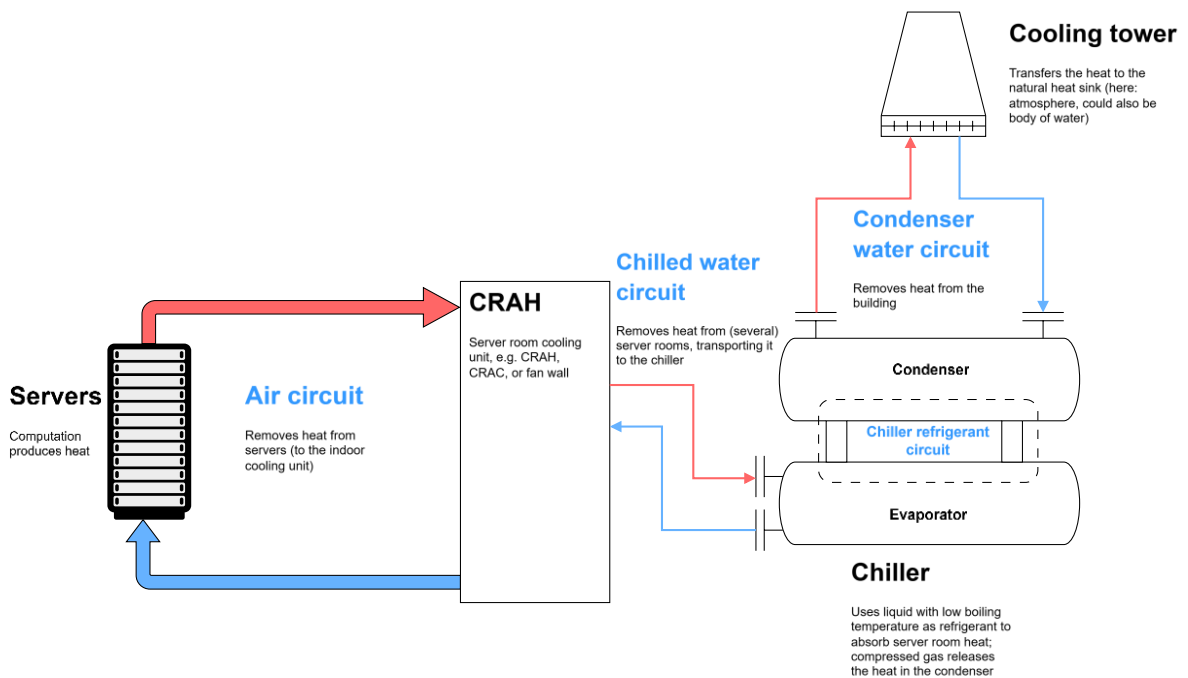

Figure 1: A reference schematic of heat removal in modern data centres, comprising four cooling circuits which sequentially remove the heat away from the servers, the server room, and finally the data centre itself to the environmental heat sink.

**Indoor cooling units**

The hot air is blown from servers into the hot aisle of the server room (Patel *et al.*, 2025). From there, the hot air is absorbed by indoor cooling units, which typically cool it down via cooling coils, supplying the cooled air back to the cold aisles of the server room. These indoor cooling units come in several flavours, such as (Patel *et al.*, 2025):

- CRAC ("computer room air conditioner"), traditional (albeit powerful, up to 100 kW each) air conditioning units paired with a condenser unit outside the server room.

- CRAH ("computer room air handler"), similar to CRAC, but part of a centralised closed cooling circuit with more extensive piping and pumping systems, and higher efficiency than CRAC systems.

- Fan walls, which typically have a higher capacity than CRAHs and remove the need for a raised floor in the server room, as the airflow is horizontal and no longer vertical.

- Rear-door heat exchanger (RDHx), placed within individual racks, it can be conceptually conceived as an in-rack CRAH; it can partially or completely eliminate the need for room-level airflow management.

Although there are important differences between these different indoor cooling systems, what they all have in common is that they absorb hot air from the server room, cool it down and supply the cooled air back to the cold aisles to be used again for the cooling of servers. To this end, the indoor cooling units use cooling coils with cold water or another liquid, which warms up while extracting heat from the air that originated in the hot aisles. Whether the cold air came from a raised floor (typical for CRACs and CRAHs) or horizontally (fan walls), the hot air is then absorbed above the servers through a chimney-like design. An exception are RDHxs, which cool the air directly at its source.

The heat absorbed by the indoor cooling units (either by warming facility water in CRAHs/RDHx or through refrigerant in CRACs) must ultimately be transferred to the outside environment. This happens via either of two main means (each of which have two main sub-flavours, as discussed below): *chillers* and *economisers*. We discuss them sequentially.

**Chillers**

Unless specific environmental conditions are met (as discussed shortly), chillers are the norm. A chiller is a machine that removes heat from a liquid via a refrigeration cycle. The main components of a chiller are the evaporator, compressor, condenser, and expansion valve (of which Figure 1 shows the most important ones, the evaporator and the condenser); see (Alpine Intel, 2025) for more details. In a nutshell, they work as follows:

1. **Evaporator:** The evaporator is a heat exchanger where the heat from the process water is transferred to the chiller's refrigerant. This refrigerant is a special fluid with a low boiling point. As the refrigerant absorbs heat from the process water, it evaporates and turns into vapour.

2. **Compressor:** The compressor takes the low-pressure refrigerant gas from the evaporator and compresses it, which increases both its pressure and temperature.

3. **Condenser:** The condenser is another heat exchanger where the heat from the hot, high-pressure refrigerant gas is transferred to a separate cooling circuit, typically water, which will ultimately transfer the heat to a natural heat sink (e.g., via cooling towers). This transfer is typically done by using air or water to cool the condenser coils. As the refrigerant releases heat, it condenses back into a liquid.

4. **Expansion Valve:** The expansion valve is a metering device that controls the flow of liquid refrigerant back into the evaporator. The temperature of the refrigerant, already lowered in the condenser, is further lowered through expansion, and the cycle can restart in the evaporator.

Two main types of chillers exist:

A. *Water-cooled chillers* are shown in the reference design from Figure 1. As discussed, they deploy an internal refrigerant circuit to perform two heat transfers: From the server room to the refrigerant in the evaporator, and from the refrigerant to the final cooling circuit (which dumps the heat to the ultimate heat sink) in the condenser.

B. *Air-cooled chillers* also perform a refrigeration cycle involving a refrigerant. The main difference is that the condenser is located outdoors, and the refrigerant transfers heat directly to the outside air via fans. The air-cooled chiller unit thus essentially combines the functions of a chiller and a cooling tower.

**Direct environmental heat sinks / Economisers**

Chillers, while highly efficient, are a main source of energy consumption in the cooling of data centres. Given suitable environmental conditions, they can be bypassed and the heat from the indoor cooling units can be dumped directly into environmental heat sinks. These can be either bodies of water or the atmosphere:

A. Given the availability of a close-by water body such as a lake, river or the sea, its water (which has typically a temperature lower than 20 degrees centigrade) can be used to cool down the warm liquid from the indoor cooling units. This happens in closed circuits; the two liquids do not mix, only the heat being transferred.

B. In locations with a cold climate (at least for a good part of the year), the data centre can use the exterior air temperature to cool the warm liquid from the indoor cooling systems. To this end, the liquid is typically pumped through pipes directly to cooling towers, where the heat transfer to the atmosphere takes place.

Given that these heat sinks usually do not require monetary compensation, have seemingly unlimited availability, they are often referred to as **free cooling** or **economisers**, while the two flavours are often referred to as **waterside economiser** and **airside economiser**, respectively.

From an environmental perspective, the names (and especially the first one) are problematic: Given the separate circuits, the water body and the atmosphere are indeed not polluted; however, raising their temperature can have longer-term consequences both on the climate and the local ecosystems. This topic, however, will not be explored further.

For both chillers and economisers, the final heat sinks can also be artificial instead of natural. Designed to reuse the heat produced in data centres for a secondary purpose, these can be both air-based (e.g., when the warm air is used to heat up an animal farm in winter) and water-based (e.g., for warm water used in district heating). While environmentally really meaningful, technologically they do not substantially alter the heat removal circuits from the DC, so do not require a separate treatment here.

2.2.3. The complexity of heat removal and further concluding remarks

Although the presentation above has been quite schematic and did not dive into various details and technological sub-flavours, it already shows the complexity of the design of data centre cooling systems. As highlighted in blue writing in Figure 1, there are several closed-loop circuits among which heat exchange happens, and which jointly help transport the heat from the IT equipment where it is produced to the natural heat sinks outside the DC:

1. the *air circuit* that starts in the cold aisle, removes heat from the chips and out of servers into the warm aisle, ultimately reaching the server room indoor cooling units, which cool it down and send it back into servers from the cold aisle,

2. the *chilled water circuit*, which circulates within the cooling coils of the indoor cooling units, absorbing the heat from the air circuit above, and passing it either directly on to natural heat sinks via airside or waterside economisers (if available), or to the refrigerant of chillers in evaporators,

3. the chiller-internal *chiller refrigerant circuit*, which absorbs the heat from the indoor cooling units in the evaporator and passes it on to the condenser water circuit, and

4. the *condenser water circuit* of the condenser and cooling towers, which absorbs the heat of the refrigerant in the condenser and brings it to the cooling towers, where it is finally transferred to the atmosphere as natural sink.

As briefly touched upon in the text, there are several possible variations to this basic reference design. Among the most common are:

- air-cooled chillers, in which the condenser is located outdoors, combining the functionality of condenser and cooling tower,

- economisers, which bypass the chiller entirely (although airside economisers and chillers both use cooling towers to transfer the heat to the atmosphere, and the two also share similar basic hardware, they have different temperature targets. the water warmed up in condensers being warmer than that arriving directly from the indoor cooling units), and

- liquid cooling, in which the first circuit is based on liquid instead of air, and it can be combined with the second one.

Free cooling and chillers are not mutually exclusive; in fact, DCs often combine them. Some climates, for example, may allow free cooling for a couple of months per year, at night but not during the day in other periods and not at all in summer. Depending on whether free cooling is available and to which extent, chillers will be deployed to compensate. And even with waterside economisers, which are less dependent on seasonal fluctuations, there might not be enough free cooling capacity, so the two might be deployed in parallel.

Finally, from a certain power density upwards (around 50 kW per server cabinet), in-room air-based cooling systems are no longer sufficient and liquid cooling is required (ASHRAE TC9.9, 2019). This brings some important changes to the entire system and would require a separate analysis. It lies, however, outside the scope of this analysis, which focuses on traditional air-based cooling within server rooms. As most specialists do not expect pure liquid systems in future for the majority of data centres, but hybrid systems which combine air and liquid cooling (Coroamă *et al.*, 2025), the insights gathered here can inform that analysis nevertheless.

# 3   Research method

The research method deployed for this report was threefold: First, a theoretical background analysis provided the theoretical foundations for the subsequent analysis. Among several conceptual insights, the two main learnings from this phase were foundations of data centre cooling (as already presented in Section 2) and a detailed theoretical understanding of the trade-off between the energy required by the cooling infrastructure and the one of server fans, presented in Section **Error! Reference source not found.** below.

This background analysis was mainly rooted in literature research that included scientific papers but also industry sources and further non-academic sources such as forum posts. It was also based on two truly insightful interviews with data centre cooling specialists, Prof. Adrian Altenburger (Lucerne University of Applied Sciences, HSLU) and Martin Casaulta (Hewlett Packard Enterprise), respectively. Kudos to both.

The background analysis also quickly made clear that there is no consensus in the core question underlying this study: Which are the optimal server inlet temperatures in different contexts, and what do they depend on? The second deployed method was then a literature review into the arguments on both sides: It compared evidence and arguments in support of both higher server inlet temperatures, and those against these higher temperatures. During the literature review, it became clear that one metric often deployed to measure DC energy efficiency, the power usage effectiveness (PUE), is particularly ill-suited to assess optimum inlet temperatures, although it is sometimes used in this sense. The literature review is presented in Section 5.

Finally, we could partner with a colocation provider who raised inlet temperatures in several of its data centres from the widespread 24 °C towards the upper limit of the ASHRAE range of 27 °C, and for some server rooms even above. Through this partnership, we received primary data from several server rooms in two of its data centres. The measured data covered server room temperatures and corresponding IT power consumptions for both buildings. For one of the buildings, we also had access to the overall building's power consumption (i.e., the sum of IT and non-IT equipment). For obvious reasons, part of this data can only be presented anonymised, for example no absolute consumption values can be shared. Our partner, however, has been generous in the data that can be shared, so this study is able to present the absolute temperature values, the relative power changes, and all correlations that could be found. Section 6 presents these results.

# 4   The trade-off between infrastructure and server fan energy consumption

## 4.1 How raising the server room temperature can yield energy savings

Operating at higher server room temperatures reduces the infrastructure energy required for cooling via two main mechanisms. They are briefly addressed in this section.

4.1.1.  Increased chiller efficiency

The efficiency of a heat pump, refrigerator, or air conditioning system (and thus of chillers as well) is usually measured via its "coefficient of performance" (COP), which is defined as the ratio of useful heating or cooling provided and the energy input / work to operate the system.[1] It can be expressed as

$$COP = \frac{|Q|}{W} \qquad (1)$$

where $Q$ is the useful heat provided or removed from the system, and $W$ is the work input to the system over the same period. Equation 1 can be alternatively also expressed as a ratio of power and not energy, i.e., as amount of heating or cooling per time (measured in Watts or some other measure such as BTU/h) and power consumption.

As heat pumps and chillers use the environment (often the air, sometimes a body of water or the underground) as heat sink or heat source, they can achieve COPs larger than one. Typical chiller COP values, for example, range from 2 – 6,[2] a higher number being obviously better and more efficient.

A simple thermodynamic analysis indicates that the higher the temperature difference between the high-temperature heat source (i.e., the chilled liquid) and the lower temperature heat sink (e.g., the outdoor air), the more efficient the chiller can operate. And the higher the server room air temperature, the higher the chilled water temperature can obviously also be.

4.1.2.  Increased use of economisers / free cooling

Significant energy saving potential also stems from increasing the usage of economisers. The higher the server room temperature, the higher the temperature of the liquid exiting the indoor cooling units will also be, as discussed in Section 2.2.2. A higher temperature of the liquid increases the likelihood that economisers can be used (Patterson, 2008).

Especially for airside economisers (with a typically higher temperature variability of the heat sink as compared to waterside economisers), this will in tendency increase the share of time that economisers can be used. Free cooling will thus be available in more seasons and for longer hours (Torrell, Brown and Avelar, 2016).

---
[1] See https://en.wikipedia.org/wiki/Coefficient_of_performance.
[2] See https://www.stoutenergy.me/blog/uuen3lk2lqw1x654qxd4xl9ehg19m1.

In some climates, such increase might make the installation of airside economisers meaningful in the first place. Below a certain temperature threshold, the return on investment (ROI) for economisers would be to poor, and their installation thus not meaningful (Patterson, 2008); hence, only a server room temperature increase would make their installation economically sound.

## 4.2 The connection between inlet temperature and server fans power consumption

At the same time, however, increased temperatures also tend to lead to an increase in the energy consumed by the servers themselves, mainly due to an increased usage of server fans. The current section develops this argument.

### 4.2.1. The relation between inlet temperature and transferred heat

The airflow traversing servers takes up heat from the chips' heat spreaders or sinks via *convective heat transfer*. Convection is the thermal energy transfer between a surface and a fluid. The rate of convective heat transfer is described by Newton's law of cooling (McMordie, 2012):

$$Q = h * A * (T_{hot} - T_{cold}) \tag{2}$$

where

- $Q$ is the rate of heat transfer [W],
- $h$ is the convective heat transfer coefficient [$W/(m^2 * °C)$],
- $A$ is the cross-sectional area through which the heat is flowing [$m^2$], and
- $T_{hot}$ and $T_{cold}$ are the temperatures of the two bodies

As immediately evident from Equation 2, the larger the temperature difference between the hot body (i.e., the chip) and the cold body (i.e., the cooling air), the more heat can be transferred per amount of time. Conversely, raising the server inlet temperature increases $T_{cold}$ lowers the difference $T_{hot} - T_{cold}$; all other things equal, this leads to a lower rate of heat transfer.

### 4.2.2. Compensating for higher inlet temperature by increasing the airflow

To achieve the same cooling effect, this lower temperature difference needs to be compensated otherwise. As the surface of contact between the heat spreader or heat sink of the chip and the air is constant, the only other possibility is to increase the convective heat transfer coefficient $h$. This coefficient is not a constant, but depends on various factors, including fluid properties (viscosity, thermal conductivity, density, specific heat), the flow velocity, and the geometry of the surface.

The fluid properties are unchangeable, and the chip heat sink's surface geometry are set, but the flow velocity can be changed. Higher fluid speeds generally lead to a thinner thermal boundary layer, which means there is less resistance to heat transfer between the surface and the fluid (Vaia, 2023). This reduced resistance translates to a higher value for the convective heat transfer coefficient (Cadence System Analysis, 2022), thus increasing the rate of heat transfer. A higher airflow velocity can thus compensate for the higher inlet temperature.

### 4.2.3. The relation between fan speed and power consumption

A higher airflow velocity through can be achieved by increasing the server fan speed. While airflow is directly proportional to fan speed, however, the power consumption of the fans unfortunately is not. It is, in fact, proportional to the cube of the fan speed.

This relatively counterintuitive relationship can be best understood while looking at the inverse system: Not one in which power is transformed in airflow, but one in which airflow is transformed into power, i.e., wind turbines. As for any object, the kinetic energy of the wind is computed according to the well-known equation

$$E_{kin} = \frac{m * v^2}{2} \tag{3}$$

where $m$ is the mass of the wind and $v$ its speed. As power is energy divided by time and the mass of air can also be expressed as its volume $V$ times the air density $\rho$, the power of the wind can be expressed according to Equation 4 below (Snurr and Freude, 2024):

$$P_{wind} = \frac{\rho * V * v^2}{2 * t} \tag{4}$$

The volume of air that passes the rotor in the time $t$ equals the wind's speed $v$ times the time $t$ times the surface area of the rotor $A$, as shown in Equation 5 below:

$$V = t * v * A \tag{5}$$

Finally, replacing the volume from Equation 5 into Equation 4, yields the final equation for the wind power

$$P_{wind} = \frac{\rho * A * v^3}{2} \tag{6}$$

As Equation 6 shows, the power of the wind is proportional to the wind speed at the power of 3. Intuitively, this can be understood as two powers come from the wind speed (as kinetic energy, and thus also power, is proportional to the square of speed) and the third power comes from the wind mass within a given time, which is also proportional to the wind speed.

The other way around, when transforming electrical energy into kinetic air energy, the relation is identical: The power needed to generate an airflow is proportional to the third power of speed of the resulting airflow (Rose and Sky, 2019); a relation known as "fan affinity laws" (Upsite Technologies, 2025). Thus, to double the fan airflow, the energy needs to grow by a factor of eight. For a less dramatic increase of 10%, for example, fan energy grows by 33%.

4.2.4. The resulting trade-off between infrastructure and server fans energy consumption

Summarising the above, a higher inlet temperature lowers the temperature difference between the electronic components of the server and the cooling air and thus reduces the rate of heat removal. To compensate for it, the only viable option (short of changing the paradigm entirely, for example through liquid cooling) is to increase the velocity of the cooling air. While the speed of the fluid (which, for servers in DCs, is air) is not explicit in Equation 2, it is influencing the rate of heat transfer nevertheless by increasing the convective heat transfer coefficient $h$.

This increased airflow could be achieved through higher speed of the fans in the fan wall (if the server room is equipped with them), which would lead to a trade-off between two types of infrastructure energy consumption: of cooling down the inlet air and to power the large server room fans. In the context of this study, we do not explore this path further, which is interesting but limited to a specific type of DC architecture.

Instead, we focus on the more widespread trade-off, i.e., the one appearing between infrastructure cooling energy and server fans when increased server fan speed is needed to generate a higher airflow. As discussed above, the increase of the airflow (which compensates for a higher inlet temperature) scales linearly with fan speed. The energy consumed by server fans, however, scales with the third power of the fan speed. This is an inherent trade-off between DC infrastructure energy consumption and server fans energy consumption.

While the increase in server fan power consumption can to some extent be mitigated, e.g. through smart fan management, this trade-off remains a fundamental limitation to how high the server inlet temperature can be raised (next to other constraints such as the safe operating temperatures for the hardware or safe working conditions for DC technicians). It also represents the starting point for the subsequent search for the optimal balance within this trade-off.

# 5 Literature review

As one blog post from Data Center Dynamics points out while discussing the temperature raise announced in 2002 by a large colocation provider, Equinix, and the pushback it faced from it customers: "All Equinix has promised to do is to make an attempt to push temperatures up towards 27°C – the target which ASHRAE set 14 years ago, and which it already recommends can be exceeded" (Judge, 2023). And it is indeed quite surprising that 17 years (in the meanwhile) after ASHRAE first expanded 2008 its recommendation from the previous range 20 – 25 °C to the current range 18 – 27 °C (Beaty and Lintner, 2008), the higher end of this range still represents the exception among colocation DCs (Judge, 2023).

To shed light on possible root causes for this high inertia, the current section presents the results of the literature review. It starts by presenting the general evidence that supports higher inlet temperatures (Section 5.1) as well as specific evidence from hyperscale DC operators (Section 5.2). It then presents the counterarguments brought by the sceptical voices (Section 5.3) as well as the inadequacy of the PUE metric for this discussion (Section 5.4). Finally, it discusses the two jointly (Section 5.5), the optimisation criterion for this discussion being the overall data centre energy consumption.

## 5.1 Evidence supporting higher server inlet temperatures

There are indications in the literature that operating at the high end or above the ASHRAE recommended temperature range (18 – 27 °C) might reduce the total power use of a data centre. Section 4.1.1, for example, discussed how a temperature increase leads to improved chiller efficiency. Quantitatively, and depending on the type of chiller and further parameters, this has been measured as a COP improvement of 1 – 2.5% for each degree Fahrenheit (about 0.56 °C) of chilled water temperature increase (Patterson, 2008). Unfortunately, it is not possible to put this number in relation to DC infrastructure energy savings, and much less to balance it against the likely increase in server energy consumption.

A stronger argument is put forward by (El-Sayed *et al.*, 2012) through an experiment including various types of workloads performed at different temperatures while at the same time directly measuring the server's power consumption. The results show that until some 30 °C, the server's energy consumption remained fairly constant for each of the workloads. Starting from about 30 °C, however, the energy consumption of the servers grew abruptly, around 40 °C reaching its peak of roughly 50% higher overall energy consumption (i.e., for computing and fans) as compared to the temperatures below 30 °C. As around 40 °C the server fans are at their maximum capacity, and further temperature increases to 60 °C did not bring any additional power consumption increase was interpreted as evidence that the increases between 30 – 40 °C were indeed due to the increased power consumption of the fans and not of the servers themselves (e.g., due to leakage power).

The SwissEnergy programme led by the Swiss Federal Office of Energy (SFOE) recommends the upper end of the ASHRAE recommended range for Switzerland. The title itself states: "Do not cool your servers below 27 °C" (EnergieSchweiz, 2022); albeit not beyond that either. A survey conducted 2021 on behalf of SwissEnergy as well, found that 25% of data centre operators did indeed operate their server rooms at 27 °C or more. Among the data centres of large companies ("Segment B" of the study), however, this proportion was negligible (Jakob, Müller and Altenburger, 2021).

## 5.2 Evidence from hyperscale operators supporting higher temperatures

Operators of hyperscale data centres are also a good indication: As opposed to operators of colocation DCs, hyperscale operators have control over both the infrastructure and the servers, and must shoulder both energy consumptions. They thus have an inherent interest to minimise the sum of the two, i.e., the overall energy consumption of the DC.

In the early 2000s, evidence from hyperscale operators started to mount that raising the inlet temperature does indeed yield energy – and consequently, financial – savings. Sun Microsystems argued that for every degree (Fahrenheit) of temperature raise, 4% of energy savings can be achieved, while Microsoft could save substantial sums by raising the temperatures by 2 – 4 °F, i.e., 1.1 – 2.2 °C (Miller, 2008). These early analyses, however, are less relevant today; back then, most data centres were operated between 20 – 22 °C (Swinhoe, 2024), which – according to current knowledge – was unnecessarily conservative. Going up a few degrees to today's standard 24 – 25 °C meant picking the ripe fruits and does not necessarily have any validity for a further increase in temperature.

More relevant for today is another experiment from the period, in which Intel ran 450 servers with airside economisation only in a relatively warm environment, with outside temperatures of 18 – 33 °C during the test. Although the server inlet temperatures went as high as 33.3°C, there were still substantial savings of 67% as compared to servers in a traditional air-conditioned environment (Miller, 2008; Judge, 2023). The savings by not using chillers were substantially higher than the additional consumption of server fans. At the same time, there was no consistent increase in equipment failure rates during the 10-month long test (Miller, 2008).

The most compelling evidence in favour of higher server inlet temperatures, however, is that today, the major hyperscale DC operators operate their servers at temperatures around the higher end of the ASHRAE recommended range for class A1 – A4 servers (i.e., 18 – 27 °C) or slightly above it. Table 1 summarises the available data.

Table 1: Important hyperscale data centre operators and their recently reported server inlet temperatures.

| Hyperscale operator | Inlet temperature [°C] | Source |
| --- | --- | --- |
| Google / Alphabet | 26.6 | (Judge, 2023; Swinhoe, 2024) |
| Microsoft | 27 | (Swinhoe, 2024) |
| Facebook / Meta | 29.4 | (Judge, 2023; Swinhoe, 2024) |

## 5.3 Evidence against higher server inlet temperatures

An experiment performed at the Research Institute of Sweden (RISE) and published in 2020 holistically considered the overall energy consumption of servers, server fans, and cooling infrastructure jointly. Depending on the average CPU load, the experiment revealed ranges of CPU temperatures for what the authors name the "holistic sweet spot" of energy consumption as follows (Sarkinen *et al.*, 2020):

- 40-50 °C for 0% load (i.e., idle),
- 52.5 – 57.5 °C for 50% load, and
- 60-65°C for 100% load.

These temperatures are CPU temperatures and not server inlet temperatures. The two correlate, of course. For inlet temperatures, the paper only looks at three discrete temperatures: 16, 22, and 28 °C. Compared to the default 22 °C and depending on the load, the experiment shows an energy consumption increase of 0.9 – 1.4% when raising the inlet temperature to 28 °C, and a decrease of 0.6 – 1.2% when decreasing this temperature to 16 °C (Sarkinen *et al.*, 2020). This does not mean that 16 °C is necessarily the sweet spot for server inlet temperatures, but that it is most likely somewhere in the interval 16 – 22 °C, much lower than most of the literature discusses.

An article supporting higher temperatures (Judge, 2023) cites Jon Summers – one of the authors of the paper describing the RISE experiment – as dissenting voice. Based on several experiments, he is cited as: "Running at hotter temperatures can create completely illusory benefits, says Summers: 'With increased supply temperatures we do see an increased overall energy consumption even though the PUE drops'." And further with the rhetorical question: "Other than less energy consumed by the DC cooling equipment resulting in a lower ISO PUE, what are the reasons for pushing up air-supply temperatures?" (Judge, 2023). Why the PUE metric is ill-suited to holistically capture the energy consumption of a data centre is discussed in Section 5.4 below.

In an interview performed for this study, Martin Casaulta (HPE), a professional with over 30 years of IT and data centre experience, largely supports this view. According to internal assessments of HPE which included direct measurements of the server energy and modelling of the infrastructure energy, "when the inlet temperature goes up, then the server draws more. And so for our servers in particular, these tests have shown that it does not really make sense to go above 26 degrees in terms of energy, because the gain in the cold aisle is then consumed by the servers drawing more because they simply have to cool themselves more" (Casaulta, 2025).

On a side note, Martin Casaulta further argued that not raising the inlet temperatures higher is also a security feature in case of cooling dropouts. By default, HPE servers support inlet temperatures of up to 35 °C. If the cooling system encounters an issue, running them at 25-26 degrees allows a buffer of about 10 degrees for a solution to be found before the servers need to be switched off (or rather, switch themselves off via self-protection mechanisms). If ran towards 30 °C, this buffer would be reduced by about 50%.

Some criticism also aims at early extrapolations from commercial buildings to data centres when it comes to the relation between indoor temperature and cooling energy consumption. For general commercial buildings, a rule-of-thumb states that for every one degree Fahrenheit of indoor temperature raise up to 2% air conditioning energy van be saved (Patterson, 2008). This stands to reason, as the energy to remove heat from the building is proportional to the temperature differential between the outside environmental temperature and the indoor ambient one. A substantial share of the building's total heat gain is external (i.e., the heat transfer through the building's envelope as opposed to the internal gain, caused by the heat produced by people and devices inside the building). As the indoor temperature approaches the outside one, this external heat gain decreases significantly (Patterson, 2008). The particular context of general commercial buildings, however, does not apply to data centres, as data centre heat load is almost entirely internal. Nevertheless, in the early days of DCs, this rule of thumb has been wrongly transferred to them as well, and some general impression might still be lingering.

## 5.4  The inadequacy of the PUE metric to discuss server inlet temperatures

Power usage effectiveness (PUE) has long been a de-facto standard in the data centre industry to measure the efficiency of operation. It is defined according to Equation 7:

$$PUE = \frac{P_{DC}}{P_{IT}} = \frac{P_{IT} + P_{non-IT}}{P_{IT}} = 1 + \frac{P_{non-IT}}{P_{IT}} \qquad (7)$$

as the total DC load divided by the IT (i.e., server) load. The PUE is always larger or equal 1, and the ideal PUE of 1 would mean that no power whatsoever was wasted on non-IT loads. Among non-IT power consumption are – in what is typically a growing order of importance – the (usually negligible) lighting and power used in the adjacent offices, the more important power transformation losses inside the DC, and the power consumed by the DC's uninterruptable power supply (UPS), By far the dominating non-IT consumption is typically the cooling-related infrastructure consumption.

While the PUE has been historically very helpful to determine the ballpark of DC efficiency, it also has several known flaws, e.g. the fact that it does not account for the possible reuse of waste heat (in district heating, for example) or that it is a relative effectiveness and not an efficiency metric; it does not say anything about the efficiency of the IT equipment itself.

In the context of this study, however, another drawback of the PUE is relevant. When looking at the last form of expressing the PUE from Equation 7, the infrastructure energy is in the numerator part of the fraction (and, as discussed above, it represents the most important part of it), while the server fans are indistinguishable from the IT consumption of the servers and are thus (semantically not very fortunately) included in its denominator.

As discussed in Section 4.1, raising the inlet temperature in a server room will lead to a decrease of cooling energy and thus of the PUE numerator. Often, this will lead to an increase of server fan power consumption, and thus of the IT load, i.e., the denominator of the PUE. Decreased numerator and increased denominator obviously lead to an overall decrease of the fraction and thus of the PUE. But this is *just a natural consequence of how the PUE is being defined and means nothing whatsoever for the overall energy consumption of the DC*.

Purely mathematically, the PUE will always decrease after a temperature increase. The overall consumption of the DC, meanwhile, might have decreased as well (when infrastructure savings are larger than the additional server fan consumption), stayed constant (when the two are balanced) or increased (when the latter overcompensates for the former). The PUE is simply not able to answer this question.

## 5.5 Discussion

Arguably the most serious evidence in favour of rather high inlet temperatures is that hyperscalers operate their DCs in the 26 – 29 °C window, as argued in Section 5.2 above. And hyperscale operators are in the best position to squeeze the last bit of efficiency out of their data centres: Having control over both infrastructure and servers, they are in the position to optimise them jointly. Paying for both electricity bills, they also have a core interest in doing so. Owning and operating the largest data centres in existence for decades already, with vast experience, highly qualified experts and huge budgets at their disposal, puts them in the best position to make these calls.

Extrapolating from hyperscale to colocation operators, however, should be done cautiously. For both cost, control, and optimisation reasons, hyperscalers typically do not buy off-the-shelf servers, but design their own solutions (Patel *et al.*, 2025) that are finely tuned for smooth and energy efficient interaction with their data centres. By contrast, colocation operators have no control over the servers in their data centres. Being owned by the colocation company's clients, these servers will not only not be fine-tuned to the DC infrastructure, but they will also be quite heterogeneous. The server fan power consumption will thus almost certainly start growing sooner and/or more abruptly than for those in hyperscale DCs, likely sooner reaching the threshold from which the growth in fan consumption is larger than the infrastructure energy savings.

Given these considerations and also the interview cited earlier (Casaulta, 2025), no definitive answer is possible. There are several converging arguments, however, that would indicate a sweet spot for hyperscale DCs around upper end of the ASHRAE recommended range (i.e., 27 °C) or slightly above, and a slightly lower – but still towards the upper ASHRAE range – ideal inlet temperature for colocation DCs around 25 – 26 °C.

# 6  Colocation case study

For the case study, we partnered with a colocation provider in Switzerland. Through this partnership, we had access to primary data from several server rooms in two of the provider's data centres. For both data centres, the data covers server room inlet temperatures and corresponding IT power consumption. This allows an analysis of the variation in server energy consumption as a consequence of temperature variations.

For one of the sites, we additionally also had access to the overall building's power consumption (i.e., the sum of IT and non-IT equipment). This enabled a direct analysis of our main topic of interest: the variation of the overall energy consumption (i.e., sum of servers including their fans, and of the cooling infrastructure) as a consequence of a variation in the server inlet temperatures.

For business sensitivity reasons, part of this data cannot be disclosed. In particular, the absolute energy consumption values cannot be publicly shared. For this dependent variable, relative changes will be presented, which actually do capture all the relevant information. The drawback of not sharing the absolute values is not one of result significance, but one of transparency and reproducibility. For the independent variable (i.e., server inlet temperatures), on the other hand, the study is able to present the absolute values.

The analysis will be presented in two steps: First, based on data from one of the two data centres, Section 6.1 analyses the correlation between inlet temperature and server consumption on server room level. Secondly, and based on data from the second data centre, Section **Error! Reference source not found.** sets out to explore the correlation between server inlet temperatures and overall energy consumption.

## 6.1 Correlation between inlet temperature and server consumption

From the study partner's first data centre, this study had access to data from 11 server rooms over a period of 2 years, 1 July 2021 – 1 July 2023. All data consisted of direct measurements, both cold aisle temperatures as well as room consumption data. The power consumption is measured at rack level. The cold aisle temperature is measured via several sensors in each server room.

The data from these physical sensors were aggregated to virtual room-wide sensors as follows: The data from the individual temperature sensors in a room (typically varying only by small amounts) were aggregated to an average temperature, while the consumption data were summed up to a room-wide server power consumption. The correlation analysis presented below used the data of these virtual sensors, analysing the correlation between the room-average server inlet temperature and the room-wide (i.e., total) power consumption.

In this historic data, sampling frequencies of temperatures and power consumption were different from each other, and distinct among the individual rooms. For power consumption, sampling values included 4-6 measurements per hour (i.e., one every 10-15 minutes). Temperature was measured more often, typically every 4-6 minutes. Additionally, the timestamps of the two typically did not coincide, i.e., there was no reoccurring time (such as the full hour) at which both temperature and consumption to be measured. A correlation analysis, however, requires the same timestamps and frequency. Among the existing options (nearest neighbour, time windowing, etc), a straightforward downsampling to hourly frequency with forward fill was chosen.

### 6.1.1. Overall analysis pipeline

Data preparation, correlation analysis, and overall statistical analysis were performed using an 8-step pipeline, presented in Table 2. Thereby,

- steps 1 – 2 perform data cleaning and preparation for the subsequent analysis,
- steps 3 – 4 are the actual per-server room analyses,
- steps 5 and 6 contain the entire pipeline for one server room and the entire building, respectively, while
- steps 7 and 8 contain the subsequent overall statistical analyses.

To test for both a linear but also a more general monotonic (i.e, consistently up or down but not necessarily linear) correlation, the analysis included both Pearson and Spearman correlation tests. Pearson is a correlation coefficient that quantifies the strength and direction of a liner correlation between two variables. Spearman, on the other hand, shows the strength and direction of any monotonic relationship. If the relation is linear, Pearson is the more precise and conclusive estimator with higher statistical power. Spearman, which converts the data to ranks, is more robust to outliers and thus better-suited for noisy, outlier-heavy data. However, this typically comes at the price of a loss of sensitivity. Its p-values are generally less powerful, so a Spearman test is more likely to miss a correlation in a clean and (roughly) linear setting.

Table 2: Script pipeline of the correlation analysis between inlet temperatures and server consumption.

| Step | Purpose | Input | Process | Output |
| --- | --- | --- | --- | --- |
| 1 | Data preparation | csv files with raw temperatures and power data per room | • extract per-room temperature and consumption columns from corresponding files<br>• parse different formats for timestamps<br>• combine into one file in chronological order | File with 3 columns: timestamp, temperature and consumption, ordered chronologically |
| 2 | Resampling | File from #1 | • handle missing timestamps<br>• resample to hourly frequency<br>• ensure unique timestamps | New 3-column file, hourly sampling, values for both temperature and power for each timestamp |
| 3 | Temperature change detection | File from #2 | • use rolling window approach (12-hour window)<br>• compare before / after means to detect sustained change<br>• parametrised change threshold (default: 0.8 °C) | csv file with one row per temp. change and further supporting data (timestamp, magnitude, etc) |
| 4 | Per-room correlation analysis | File from #2 and temperature changes from #3<br><br>Desired analysis window lengths $[N_1, N_m]$ | • analysis of correlation around each temperature change point<br>• by comparing average power during window ± N days around temp. change point<br>• parametrizable for any number $[N_1, N_m]$ of windows<br>• calculates Pearson and Spearman correlations<br>• computes absolute temperature sensitivity [kW / °C] and relative sensitivity [% / °C]<br>• generates plots and summary | • one plot per combination of temperature change and analysis window length $[N_1, N_m]$<br>• one correlation file. Per same combination of temperature change and analysis window length $[1, m]$, a row recording correlations and sensitivities |
| 5 | Single room pipeline | Desired analysis window lengths $[N_1, N_m]$ | • orchestrate steps 1-4 for a single room | All results from steps 1-4 combined |
| 6 | Batch pipeline | Desired analysis window lengths $[N_1, N_m]$ | • automates pipeline for all server rooms in main directory<br>• scans directory and extracts room numbers<br>• calls step 5 for all rooms with the (same) desired analysis window lengths | All results from steps 1-4 combined, for all rooms available |
| 7 | Statistical analysis | - [none] | • combines all correlation data from each room into one file<br>• performs overall stats analysis<br>• compares correlations across window lengths (ANOVA) | • overview plots<br>• regression plots with $R^2$ and "sensitivity gradient" |
| 8 | Temperature change analysis | - [none] | • analysis of temperature change patterns | • analysis and plots of number of changes, timeline, magnitude distribution |

### 6.1.2. Temperature change detection

To keep the pipeline generic and usable for any data from any number of server rooms, an automatic temperature change detection was required. A naïve approach would be to compare the temperatures at subsequent timestamps. This approach would be very sensitive to outliers in the data, though, either due to short-term temperature fluctuations or even faulty sensors.

Instead, an approach based on rolling windows was deployed. Two adjunct rolling windows of the same length sweep the (downsampled, hourly) data. The mean temperature is computed for each window, and only when the difference between the temperatures averaged over such a longer period exceeds a certain threshold, a temperature change event is triggered. After some experimentation, a window length of 12 hours and a temperature threshold of 0.8 °C proved sufficiently robust to noise, while still sensitive enough to catch all changes.

Along the 2-year period under scrutiny, a total of 65 temperature changes were thus detected. They occurred across the entire period and in all rooms but were quite unevenly distributed both across the 11 server rooms (between 2 – 9 changes per room) and across the period (0 – 13 changes per month, with changes occurring in 18 / 24 months and peaks in August 2022 and June 2023, as shown in Figure 2).

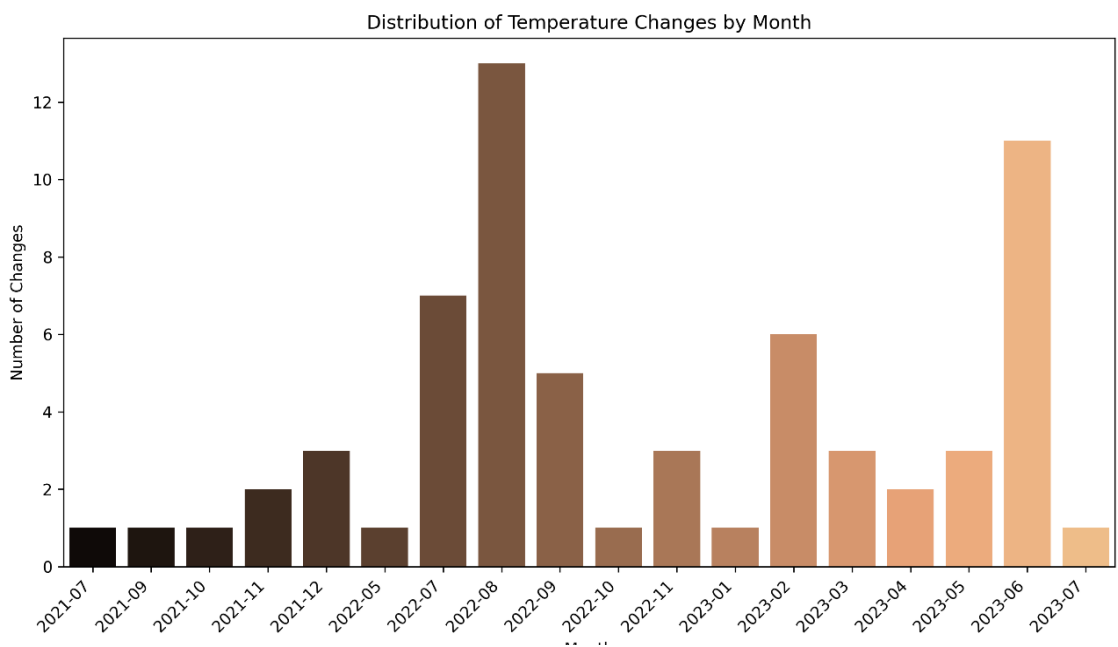

Figure 2: Distribution of temperature changes (aggregated for all 11 server rooms) along the 2 years under scrutiny. Only months with temperature changes are shown.

The accuracy of the temperature change detection was assessed by comparing the detected changes with per-room plots of the inlet temperatures along the entire period. The precision was very good, with zero false positives. There were, however, two false negatives: In one of the rooms, the provider had experimented with two half-degree temperature changes. Lowering the detection threshold from 0.8 °C to the required 0.4 °C to detect these two changes as well would, however, subject the analysis to noise and several false positives. As the effects of such small changes are likely less significant, these two half-degree changes were ignored.

### 6.1.3. Principles and methodological choices of the correlation analysis

After determining the temperature changes, the correlation analysis must perform a before-after comparison, relative to the time of each temperature change. This comparison shall reveal changes in the dependent variable (i.e., the server room power consumption) that could be a consequence of changes in the independent one (i.e., the inlet temperature).

An important methodological step was to determine how to perform this before-after analysis. Questions included:

1. Should the analysis compare two individual, discrete values or the averages of two intervals?
2. Which discrete values or intervals should be compared, i.e., how long (or for how long) before and after the temperature change?
3. What are the most important confounding factors and how can their influence be mitigated?

Addressing these questions required both contextual domain knowledge (acquired from repeated interviews and meetings both with the colocation study partner and one of its customers) and a preliminary data analysis. The insights thus gathered informed the following principles:

- *Fast response*: Changes of the energy consumption of servers (i.e., dependent variable) follow quickly after changes of the server room temperature (i.e., independent variable). Despite noise and confounding factors, both the theory (that higher temperatures must affect server cooling and thus fan reaction immediately) and the preliminary data analysis show that the new level of consumption settles in within the first hour after the temperature change. Hence, referring to question (2) above, the analysis must not wait any longer period of time for the reaction to settle in, but the *before and after moments or intervals can be directly adjacent to the temperature change*.

- *Noisy data*: Figure 3 shows the correlation between inlet temperature and server power consumption in one of the server rooms. The figure is relatively typical for all server rooms that have been under scrutiny. Although both temperature and power have been cleaned of outliers (which usually stem from one sensor not reporting data or reporting erroneous data, as discussed above), on hourly basis, there is still noise in the data. The presence of noise answers question (1) above: Individual values could be subject to noise and yield misleading results. The average values of before and after windows are thus preferred, as they minimise the impact of noise.

- *Cyclicality of compute loads (and thus server power consumption)*: Figure 3 also shows the cyclicality of the server power consumption between day and night. In Figure 4, both day-night and weekday-weekend cycles can clearly be seen. These cycles arise from the cyclicality of many of the compute loads in colocation DCs: Clients are often businesses with a stronger computing activity during business hours than at night or during the weekends (although some batch processes such as backups or complex calculations can be scheduled for these low-demand periods). Lower compute demand also induces lower server power consumption, although the correlation is not always linear (Coroamă *et al.*, 2025). Additionally, interviews with the colocation provider and one of its customers revealed that due to end-of-month, quartal, and yearly business activities, monthly and yearly cycles also exist. As partial answer to question (3) above, *this cyclicality is one of the strongest confounding factors*, and its effect on server power consumption typically much higher than the effect of a server inlet temperature change.

- *Overall server capacity*: Another confounding factor is the change in server capacity in any given server room of the colocation provider. Within their designated racks, customers can add, remove, or exchange servers as they please. There are often also free slots still available in a room that can be additionally bought by new or existing customers. While the overall compute capacity – and thus also the maximum server power consumption – are more inertial and change slower than the daily cycles mentioned above, when such changes do occur, they also influence the overall power consumption in a server room.

- *Consecutive temperature changes*: As can also be seen in Figure 4, sometimes two temperature changes follow a few days apart. They were either in the same direction (as in this instance) but also in the opposite direction. When the analysis window stretches over two or more such changes, the effects of the others confound the analysis of the change being under analysis.

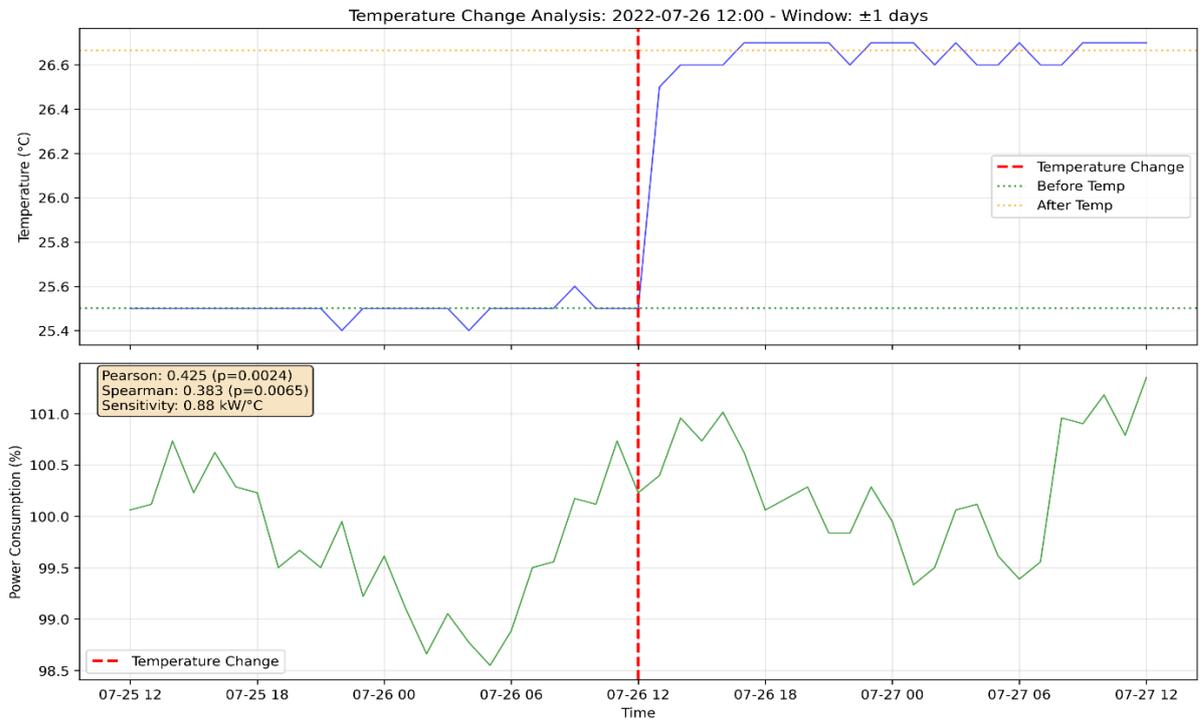

Figure 3: Correlation analysis in one of the server rooms between inlet temperature and room-wide server power consumption. The figure shows the hourly noise in the data and the daily day-night cyclicality.

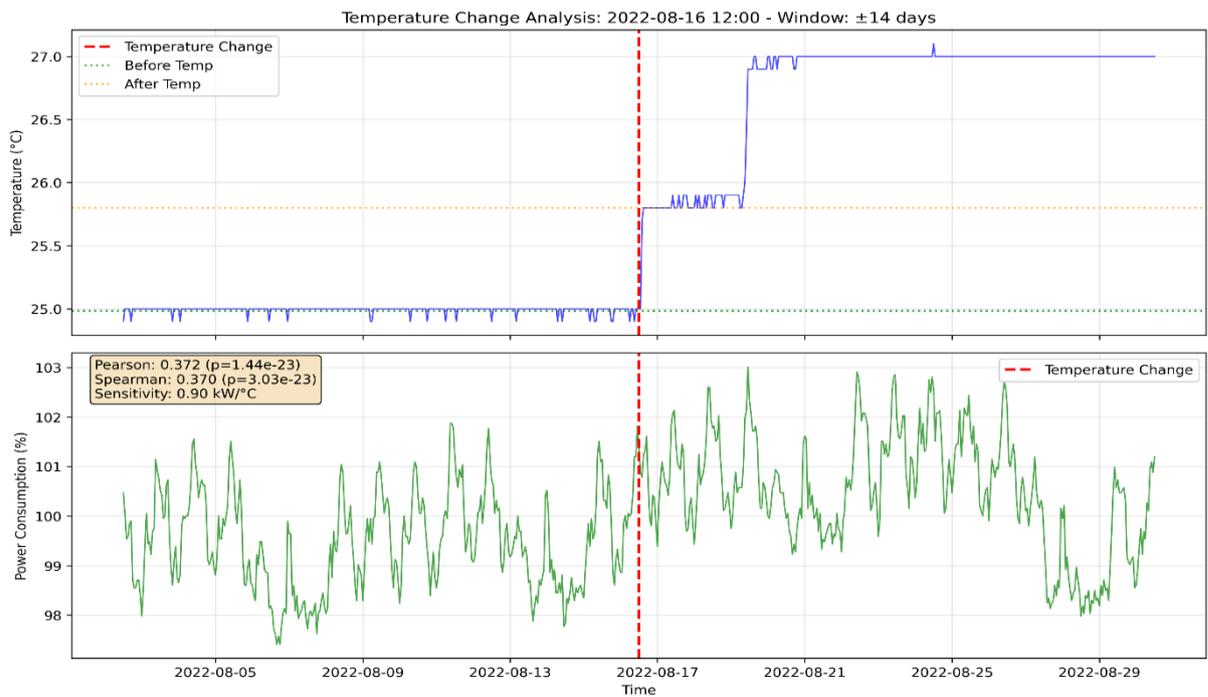

Figure 4: Correlation analysis in another server room between inlet temperature and room-wide server power consumption. The figure shows the daily cyclicality, the differences between days of the week and weekends as well as how two temperature changes can follow in relatively quick succession.

Taken together, the last three bullets show there is no ideal window length for the before-after comparison of server room power consumption. Given the hourly sampling rate, a window of a few hours (e.g., 2h or 4h) would be too short to abstract from noise and simultaneously too long to avoid the daily cycles, as can clearly be seen in Figure 3.

A minimum window length of 24h (1 day) thus appeared meaningful for the analysis. A 24h or 48h window, however, might be confounded by the weekly cycles, particularly if the temperature change takes place in the beginning or the end of the week – as shown in Figure 4 by the second change (not the one highlighted with the red dashed vertical line), which took place on a Friday.

A weekly, bi-weekly, or even monthly, window might thus be more meaningful to abstract from these short-term confounding factors. Such a long window, however, makes it more likely that the overall server capacity will have changed in the meanwhile.[3] Additionally, and specific for this case study, a longer window also increases the chances to stretch over several temperature changes with effects confounding each other.

### 6.1.4. Correlation results

Given these considerations, the study deployed windows of various sizes to control across their different confounding causes. Across all rooms, before-after windows of 1 day, 2 days, 7 days, 14 days, and 30 days were employed. Combined with the 65 temperature changes across all rooms, this yielded a total of 325 analyses. Results are presented both per window length and across all window lengths.

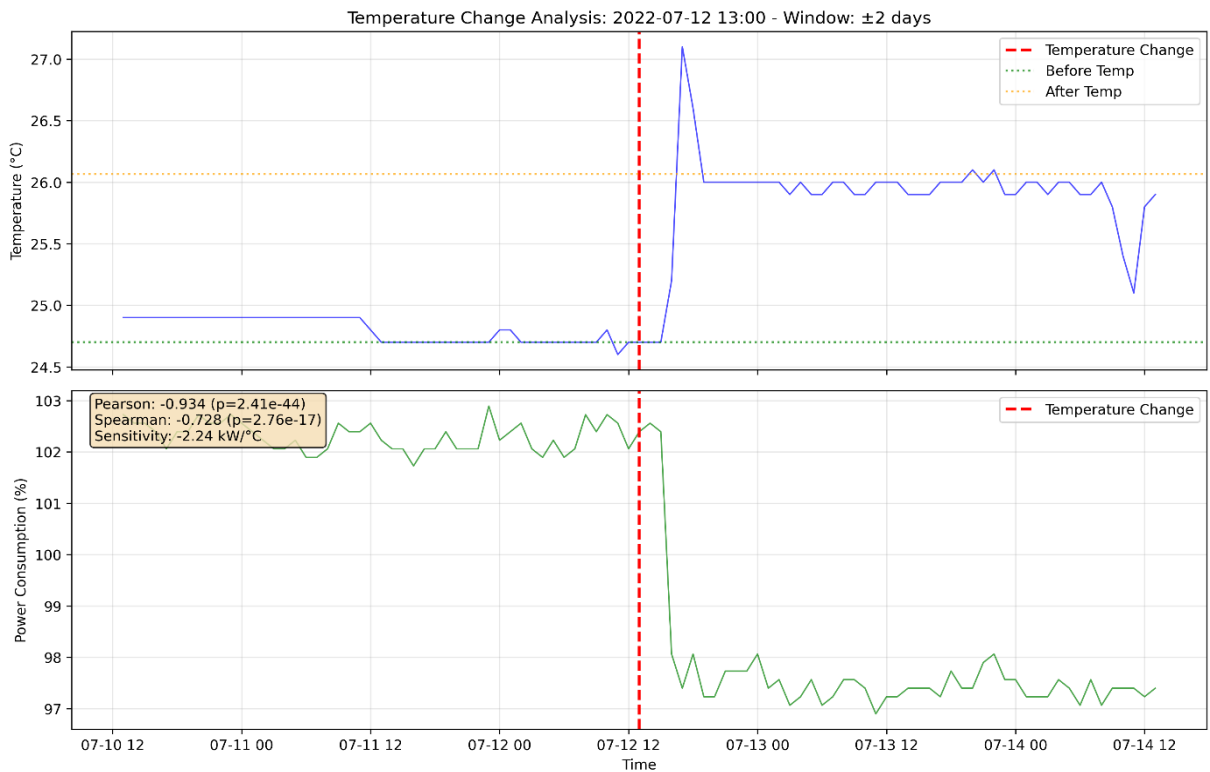

Figure 5: An example for an unexpected negative correlation between inlet temperature and server power.

This relatively large basis is important to establish statistical significance. After the theoretical analysis of the relation between inlet temperatures and server fan power consumption presented in Section 4.2, the expectation was of a positive correlation. In the two examples presented so far in Figure 3 and Figure

---
[3] Changing weather patterns are also likely to occur over weeks or months. They are not so relevant here, though, as they have little influence on the power consumption in the server room. Their influence can be very important for the energy consumption of the cooling infrastructure, as discussed in Section 0 below.

4, this was indeed the case. There are, however, also other examples in the data, showing a negative correlation between the two. Figure 5 shows a particularly striking example, in which the negative correlation is very strong and clearly observable in the graph as well. A large sample size of 65 temperature changes and a total of 325 analyses across all window lengths provides a base likely large enough for significant conclusions despite variability.

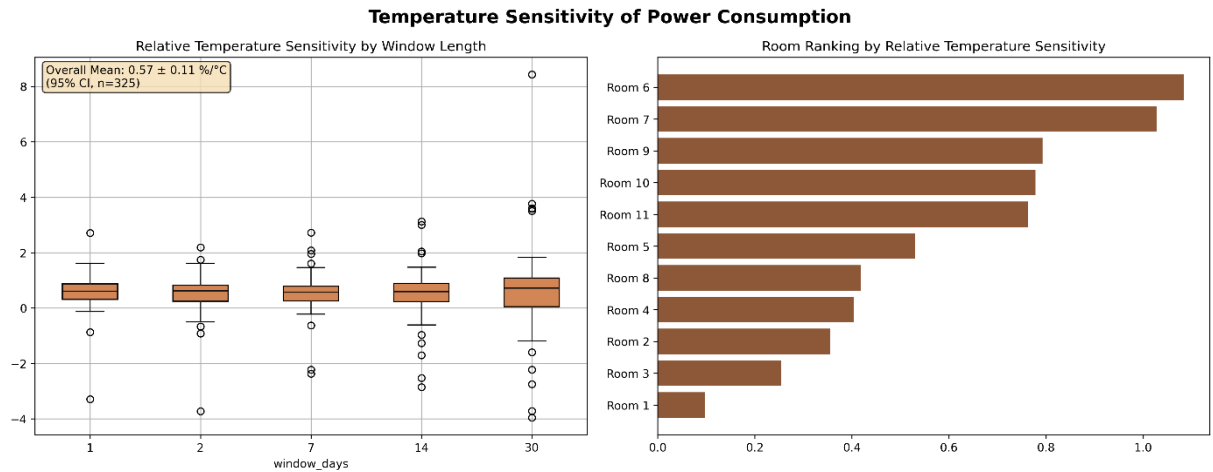

Figure 6: Aggregated temperature sensitivity of the power consumption for all rooms, distinguished by window size (left) and individual room (right).

As already shown in the legends included in Figure 3 – Figure 5, each analysis also computed the **relative temperature sensitivity**, calculated using the Pearson correlation, and expressed in [*percentage of consumption change per degree centigrade*].[4] Figure 6 summarises all 325 temperature sensitivities. The right side of the graph simply shows there is variability between the individual server rooms.

By contrast, the left side of the graph shows the sensitivities across the individual window lengths of 1, 2, 7, 12, and 30 days, respectively, each of which contains 65 values. As shown in the figure, despite some differences, the interquartile ranges of these sensitivities are all entirely above zero. For the shorter 1- and 2-day windows, the individual values are more homogeneous. For the longer windows, the results are more dispersed.

We interpret this as evidence that once the influence of the daily cycles has been controlled for (by a window that is a multiple of 24h, as used here), the long-term confounding factors (e.g., changed compute capacity or several temperature changes in short succession) have a larger influence than the remaining short-term confounds (such as noise in the data).

Across all window lengths, the **mean temperature sensitivity is 0.57 % per degree Celsius** (95% CI: ± 0.11 % / °C).

A last test was performed to analyse whether this sensitivity varies with the absolute value of the inlet temperature (whether, for example, the sensitivity – and thus the resulting variation in power consumption – is stronger for temperature changes at higher temperatures). As shown in Figure 7, the absolute temperature has a very low explanatory power, with an $R^2$ of 0.005. The temperature sensitivity thus stays virtually constant irrespective of the starting temperature, at least for the interval of 23 – 29 °C that has been analysed.

---

[4] The absolute temperature sensitivities, i.e., the absolute consumption change per temperature change (expressed in [kW/°C]), were also computed. As they depend on the overall consumption of a specific server room, they are, however, quite contextual and less relevant. They are not further addressed in this study.

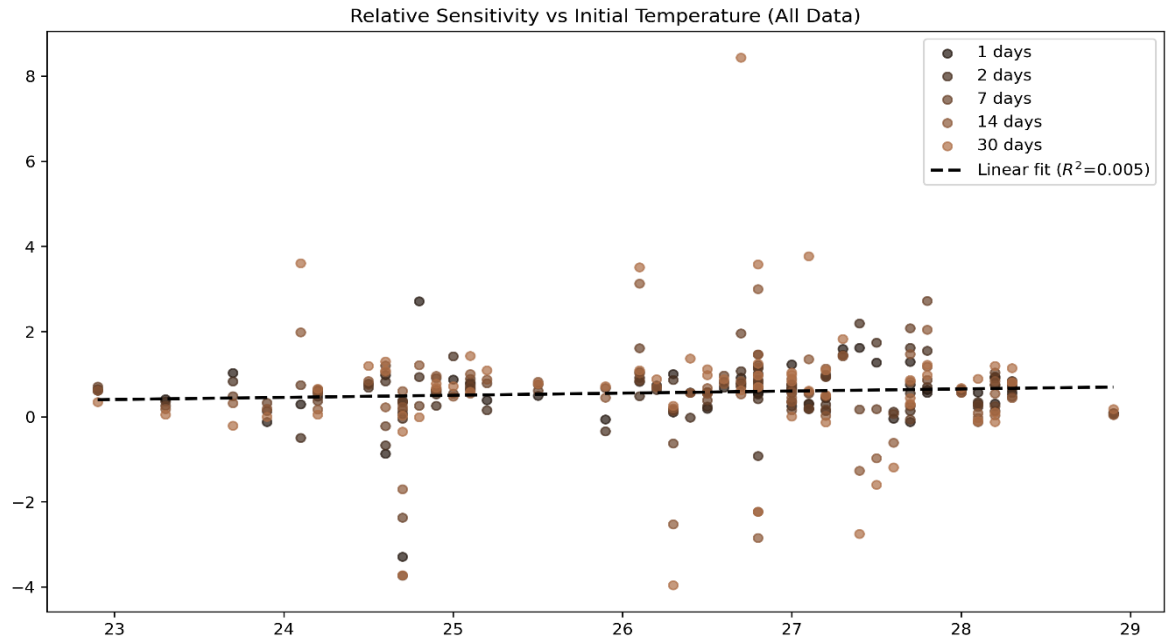

Figure 7: The dependence of the relative temperature sensitivity of power consumption on the absolute before-change absolute temperature values is very weak.

## 6.2 Correlation between inlet temperature and overall building energy consumption

Data from the second data centre of the case study partner differed from the data presented so far in several aspects:

- *Better data resolution*, with one value every minute for all server rooms. This resolution was common to both temperature and power consumption data, so that no harmonisation or downsampling of data was necessary.

- *Fewer server rooms and fewer temperature changes*: At this site, a total of 6 server rooms changed inlet temperatures along the last years: the same 4 rooms in both December 2022 and October 2023, and two further rooms in June 2025.

- *Simultaneous temperature change*: As opposed to the other DC, the temperature change in this one were coordinated and occurred simultaneously across several rooms as follows:

    o  1 December 2022: all 4 rooms changed temperature from 25 to 26 °C.

    o  31 October – 1 November 2023:
    
    - the same 4 rooms changed 26 → 30 °C at 15:30h
    - two of them changed 30 → 27 °C at 20:30h
    - the next day at 8:30h, the two went back up 27 → 30 °C
    - at 15:30h, all four rooms lowered 30 → 27 °C for a new stable level 1 degree centigrade above the former 26 °C.

    o  24 and 26 June 2025: two other rooms changed the temperature twice, 26 → 27 °C first, and 27 → 28 °C two days later.

- *Substantial share of the overall consumption*: The four rooms changing temperature simultaneously accounted for more than 40% of the DCs overall IT power consumption, and two of the

33/46

rooms were still more than 20%. These substantial shares nourished the hope that the effects of temperature changes would be noticeable not only on server room level, but also for the building cooling infrastructure, enabling a correlation analysis of temperature changes and overall building consumption.

- *Building total energy consumption*: As stated in the beginning of this chapter, for this DC data on the building's overall power consumption (i.e. IT and cooling equipment together) was available, with the same resolution of one value per minute.

- *Data for shorter specific intervals around temperature changes*: Although data was available for the entire multi-year period, given its high resolution, it was sufficient to focus on the periods one month before and after the temperature changes, analysing

    o December 2022 – January 2023,

    o October – November 2023, and

    o June-July 2025.

6.2.1. Principles and methodological choices of the analysis on server room level

Given that temperature and consumption data had both better – and especially identical – resolution than for the first DC, the only data preparation step was the removal of outliers, usually stemming from faulty sensor readings. These (few) temporary glitches were addressed by threshold analysis and interpolation between nearby genuine values.

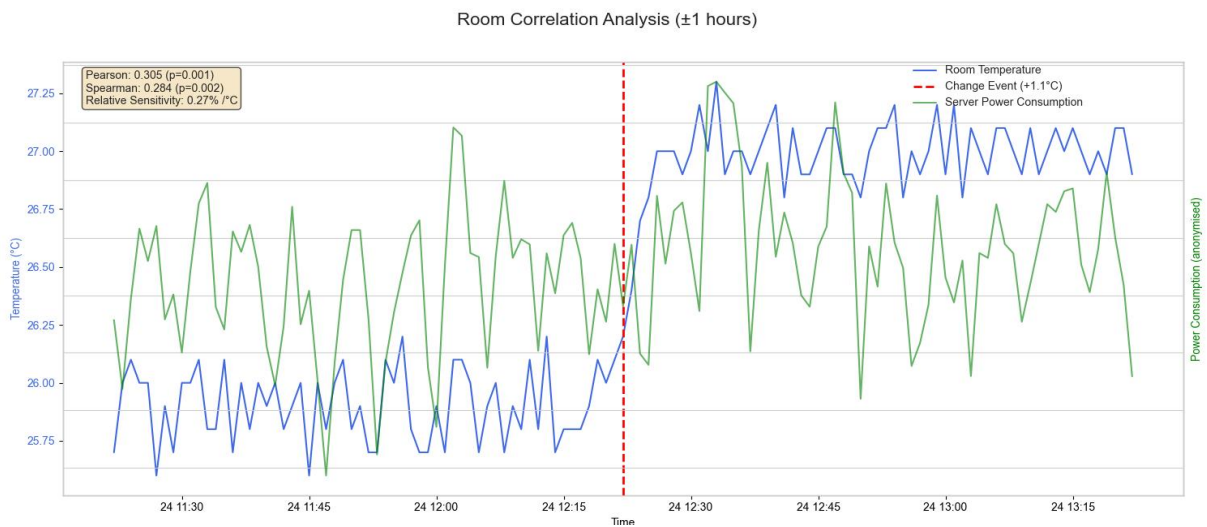

Figure 8: Example for the data from the second data centre, with 1-miute resolution for both temperature (in blue) and consumption (in green) data. To save space, for the second data centre, the two were plotted jointly.

Figure 8 shows an example for the room-level data in this second DC. A window of ± 1h already contains 60 before and 60 after values, i.e., more than double the number of values of the shortest 24h-window from the first DC with hourly resolution (after downsampling).

One hour was nevertheless chosen as the shortest window here. As Figure 8 shows, even the independent variable (i.e., the inlet temperature) takes around 15 minutes to transit to the new value. The dependent value (i.e., the consumption) likely requires longer to settle. For windows shorter than 1 hour, the spurious transition effects would thus be too strong.

At the same time, a window of 1 hour is short enough for the effects of the daily cycles to not become dominant. In this analysis with finer data resolution, how strong the influence of this confounding factor really is can be seen clearer than in the first analysis.

Figure 9 shows temperature (blue) and consumption (green) data for a window of ± 1 week (i.e., 168 h) in one of the two rooms that had 4 abrupt temperature changes in quick succession on 31 October – 1 November 2023. The two weekends with less compute demand (and thus power consumption) are obvious from the graph; without consulting a calendar, one can immediately conclude that 31 October and 1 November 2023 were a Tuesday and a Wednesday. It is also obvious that the peaks on these two days (both with the inlet temperatures raised to 30 °C as opposed to the 26 or 27 °C for the other days) are higher than the peaks of the Monday, Thursday and Friday in that week. The additional influence of the raised temperature (even one as substantial as 3-4 °C), however, is also clearly smaller than the influence of the daily and weekly server consumption cycles.

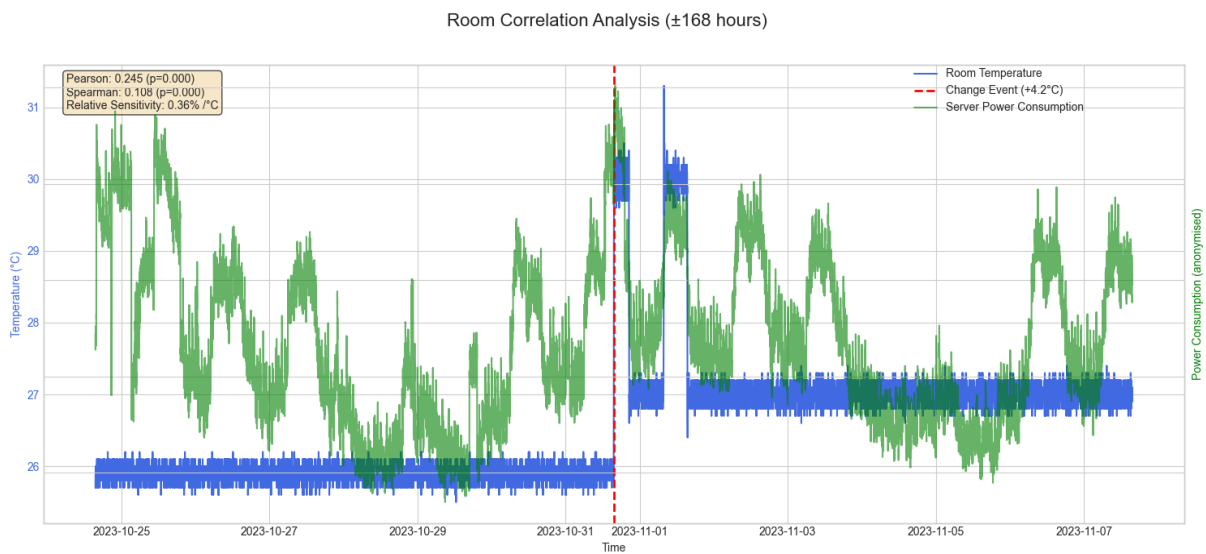

Figure 9: Temperature and consumption in one of the two rooms that had 4 abrupt temperature changes on 31 October – 1 November 2023: from 26 to 30 to 27 to 30 and finally t0 27 degrees. The daily and weekly consumption cycles (due to varying computing loads) can be clearly seen, and also that their influence is stronger than that of temperature changes, even for substantial changes are in the range of 3-4 °C.

Figure 9 further shows, however (and the numeric analysis confirms), that the values on the Tuesday and Wednesday of the previous week were also higher than the consumption of the surrounding weekdays: albeit this phenomenon does not happen in the subsequent week. The high consumption on 31 October and 1 November 2023 might thus have also been confounded by weekly cycles.

To account for these confounds, similar to the analysis of the first DC, here too different window sizes for comparison were chosen; namely 1 hour, 2 hours, 4 hours, 12 hours, 24 hours, and 168 hours (i.e., one week). The most meaningful windows have proven to be:

- 1 and 2 hours, which are long enough to already catch the effects of the temperature change and short enough to not be influenced by several subsequent temperature changes as the ones shown in Figure 9 (i.e., 4 changes within 24h). At the same time, these windows are not too strongly influenced by the daily cycles.
- 24 hours and 7 days, which avoid daily cycles, and daily and weekly cycles, respectively. However, these longer cycles risk the signal of the temperature change to be lost among the noise of other confounding factors.

Windows of 4 and 12 hours, also considered in the beginning, are not really meaningful as they are subject to the very strong daily cycles, as discussed above.

### 6.2.2. Principles and methodological choices of the analysis on building level

Next to the better data resolution, the main advantage offered by this second data centre was the availability of power consumption data for the entire DC. As discussed in the beginning of Section 6.2, several rooms also shared temperature simultaneously, accounting together for a substantial share of the DCs total IT power consumption.

The main aim of this second assessment was thus to analyse the correlations between server inlet temperatures and overall DC power consumption; specifically, to explore whether the expected infrastructure energy savings when raising inlet temperatures compensate for the (expected and from the first data centre confirmed) server-side consumption increases due to the higher fan speeds.

Similarly to the analysis on sever room level, this analysis involved Pearson and Spearman correlations for different window sizes and all temperature changes. As the temperature changes happened simultaneously and for the same start and end temperatures across several server rooms, the correlations were assessed between the building power consumption and the temperature from either of the respective server rooms. A total 7 temperature changes were analysed:

- 1 December 2022: 4 rooms 25 → 26 °C
- 31 October 2023: the same 4 rooms 26 → 30 °C
- 31 October 2023: two of these rooms 30 → 27 °C
- 1 November 2023: both rooms back 27 → 30 °C
- 1 November 2023: all 4 rooms 30 → 27 °C
- 24 June 2025: two other rooms 26 → 27 °C
- 26 June 2025: the same two rooms 27 → 28 °C

For the reasons outlined in Section 6.2.1, windows of 1, 2, 24, and 168 hours were deployed in the analysis. Figure 10 shows an example of such correlation analysis.

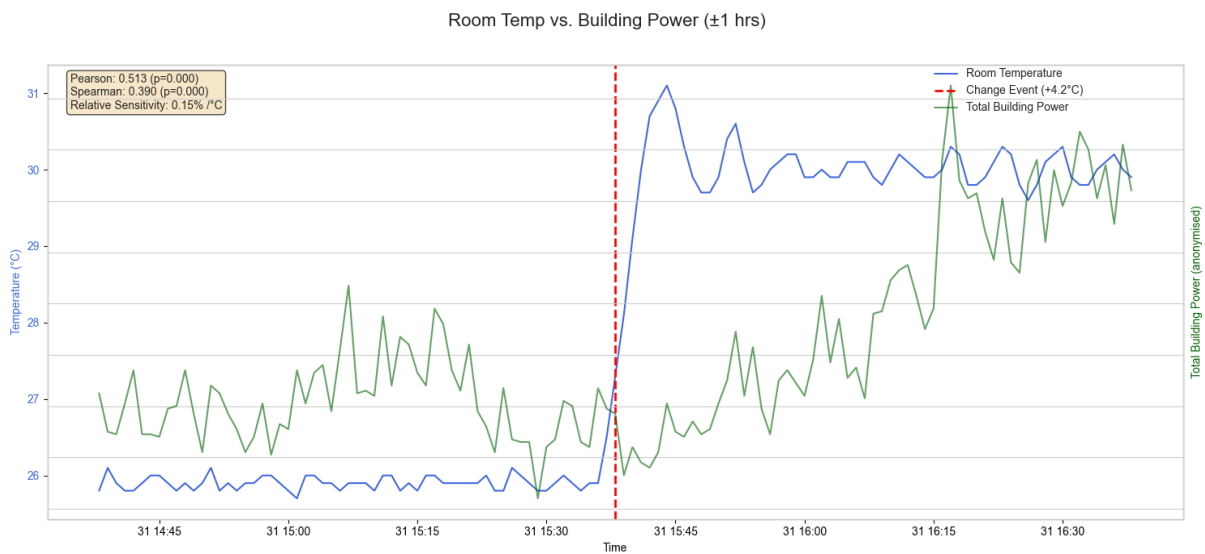

Figure 10: Example of a correlation analysis between temperature change (in 4 rooms simultaneously for this example) and the overall building power consumption. In this example, there is a strong positive correlation for a 1h-window.

Additionally, to fully abstract from both daily and weekly cycles, a further analysis was deployed on building level: A 1-hour windows comparison one week apart at precisely the same day of the week and time of that day, which could be parametrised to be e.g. 4 days before – 3 days after the change.

### 6.2.3. Correlation results on server room level

The correlation analysis on server room level largely confirms the results from the other analysis in the first data centre. There is a positive correlation between the inlet temperatures and server power consumption, most likely due to the increased server fan speed at higher temperatures. This positive correlation is most noticeable for the large temperature changes of 3 or 4 degrees performed on 31 October – 1 November 2023; Figure 11 presents a typical example.

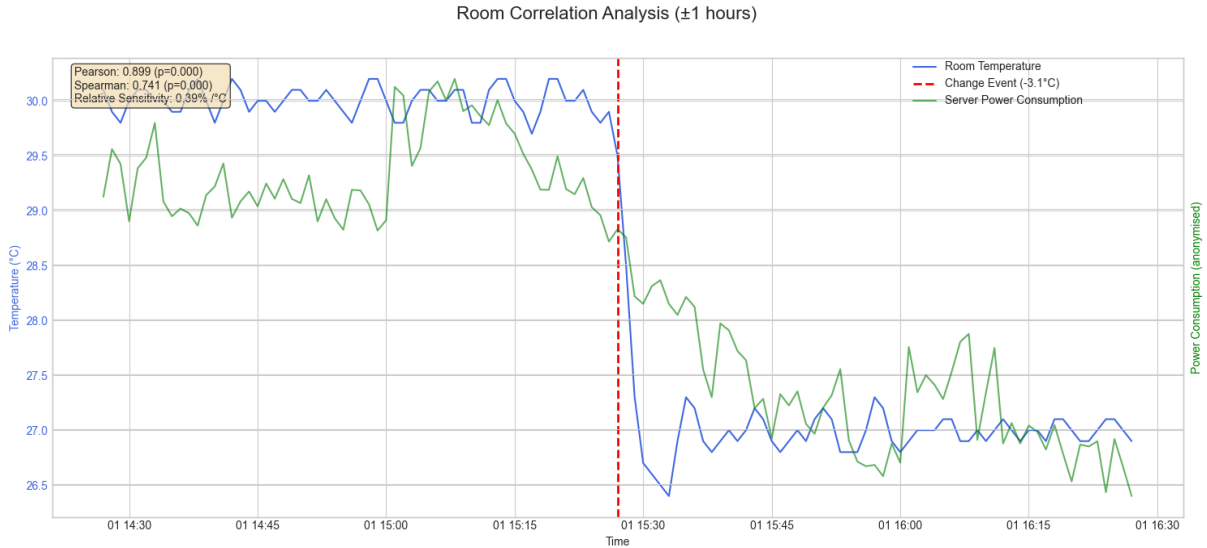

Figure 11: Example of clear server-room positive correlation between inlet temperature and server consumption for a -3 °C change in temperature and an analysis window of one hour.

This clear result can be noticed across all temperature changes and for all window lengths (i.e., 1 and 2 hours, 1 day, and 1 week), as shown in Figure 12:

- For the all-important 1-hour and 2-hour windows sizes, inlet temperatures and server power consumption are positively correlated for all 7 temperature changes.
- For each of the longer 1-day and 1-week windows, 6/7 correlations are positive and 1/7 is negative.

Thereby, as shown in Figure 13, all 1-hour, 2-hour, and 1-week correlations are significant, and also most of the 1-day correlations. Across all window lengths, the **mean temperature sensitivity is 0.34 % per degree Celsius**, which incidentally correlates very well with the 1-hour sensitivity.

Compared to the analysis of the first data centre presented in Section 6.1, the data here are likely more reliable for two reasons:

- the far superior time resolution, which data points every minute instead of every 5-15 minutes, downsampled to 1 hour (which allows analyses with short windows such as 1 hour to be significant without becoming subject to the daily consumption cycles), and
- from the seven temperature changes in the dataset, four were of 3-4 °C, thus allowing potential phenomena and correlations to emerge more clearly and distinguish themselves from noise in the data.

Given the evidence from both data centres, but especially from this second one, it can be argued that server energy consumption correlates indeed positively with the server inlet temperature, and that the temperature sensitivity of the former (dependent variable) is about 0.35 – 0.5 % / °C, where the percentages refer to the overall power/energy consumption in the server room (and not only to the server fans, which have not been analysed separately). The remaining is whether the expected cooling infrastructure savings can compensate for this increased consumption in the server rooms.

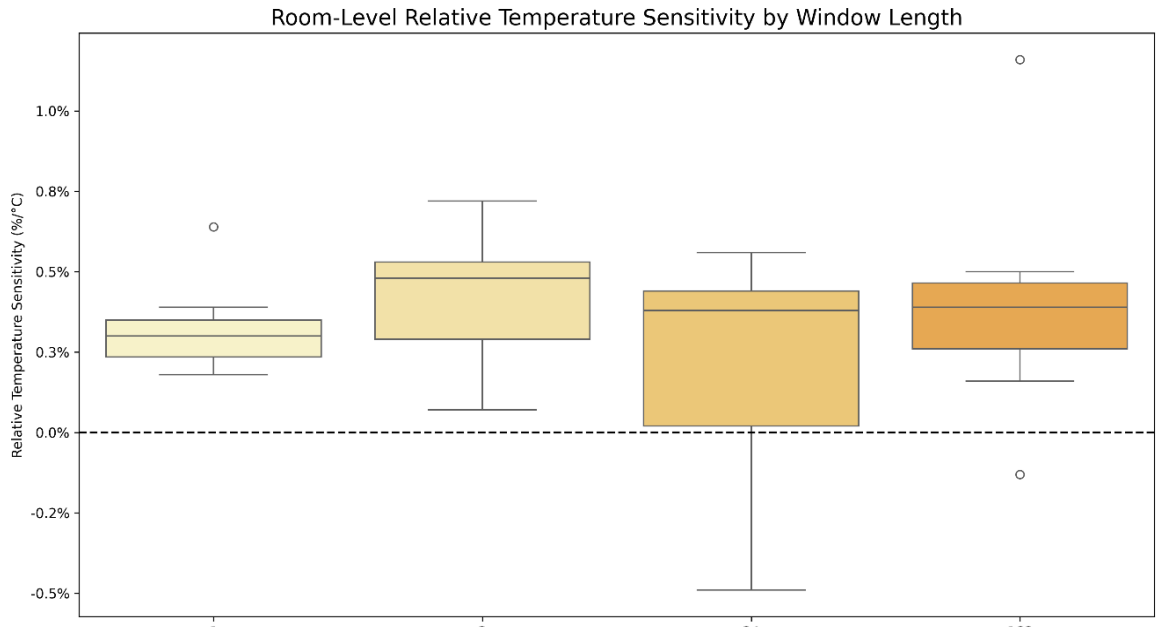

Figure 12: Aggregated temperature sensitivity of the power consumption for all rooms, by window size.

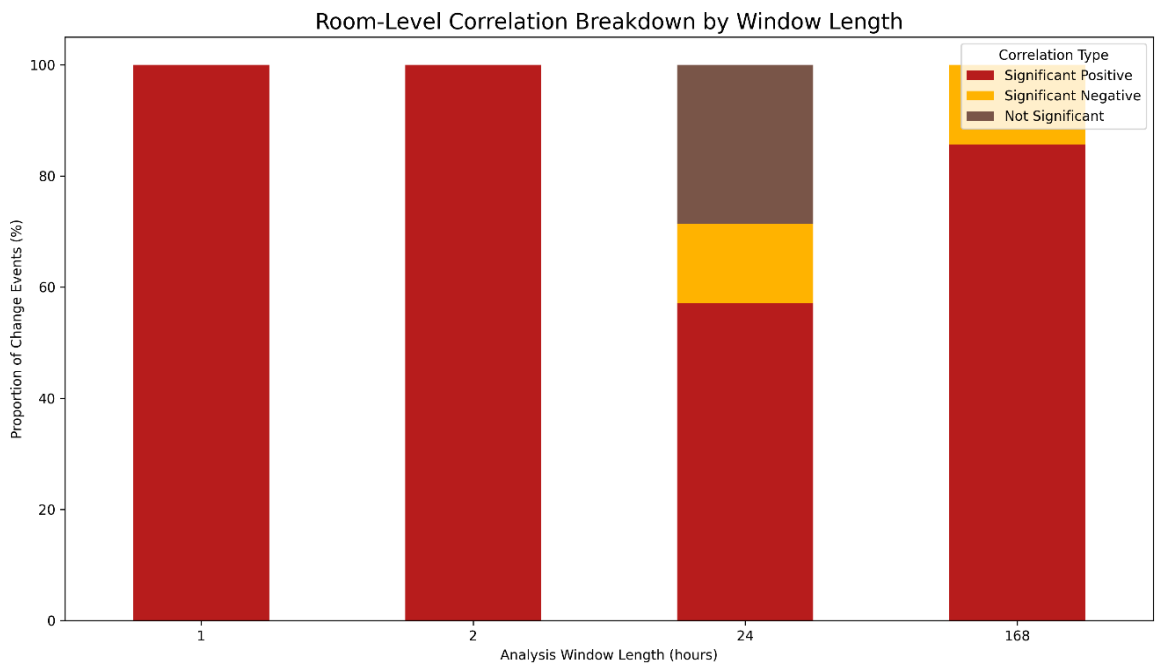

Figure 13: Direction and significance of the correlations between the changes in inlet temperatures and changes of the server room power consumption, grouped by the length of the analysis window.

### 6.2.4. Correlation results on building level

The aim of the analysis on building level was to reveal possible correlations between server inlet temperatures and overall data centre power consumption, to reveal whether. the expected infrastructure energy savings when raising inlet temperatures compensate for the server-side consumption increases due to the higher fan speeds discussed above.

As discussed in Section 6.2.2, two types of analyses were deployed: The first one, similar to the correlation analysis between inlet temperature and server energy consumption, compared before-after temperature change windows with lengths of 1, 2, 4, and 168 hours. The other analysis compared the building consumption for 1 hour on identical days of the week and times of those days. It which could be freely parametrised which days to compare, e.g. 4 days before – 3 days after the change, or 3 – 4, 5 – 2, 7 – 7, etc.

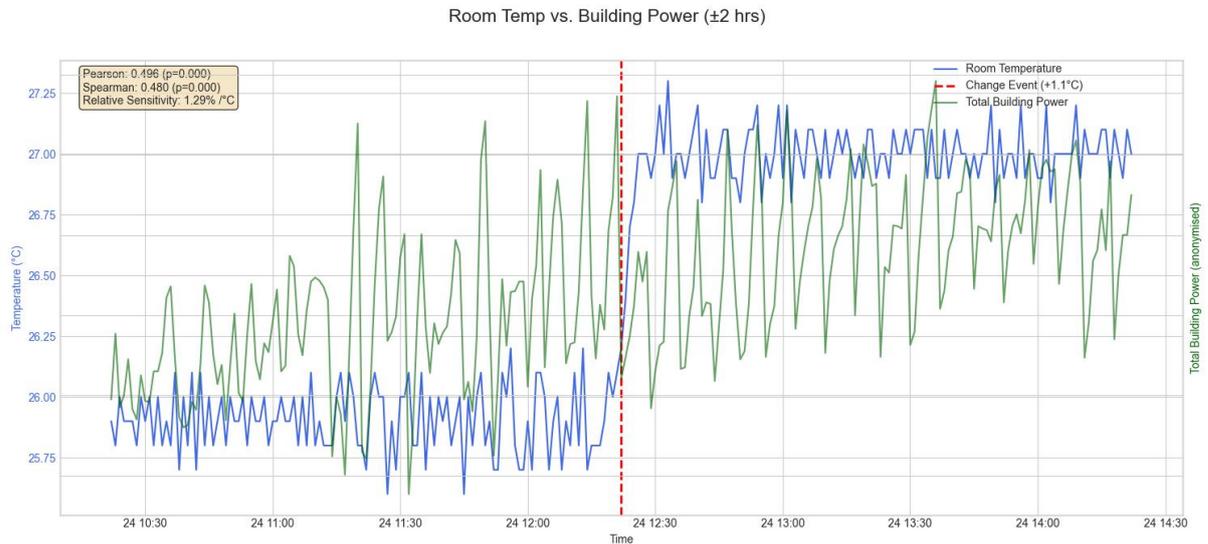

Figure 14: Temperature vs. overall building power consumption (2-hour window) for the temperature raise from 26 – 27 °C on 24 June 2005, noon.

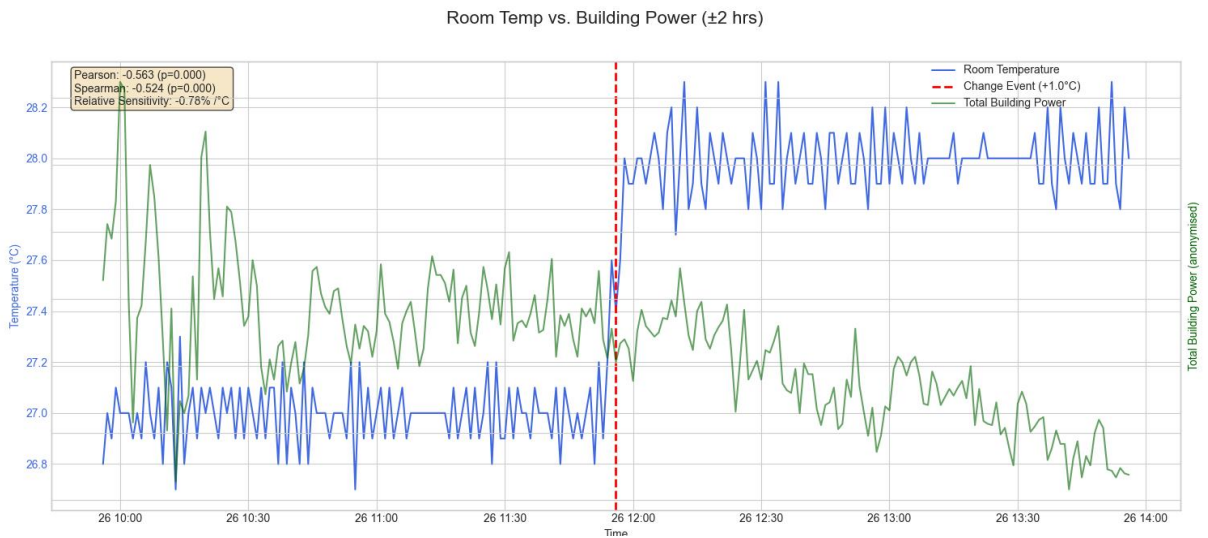

Figure 15: Temperature vs. overall building power consumption (2-hour window) for the temperature raise from 27 – 28 °C on 26 June 2005, noon.

Unfortunately, the results of the building-level analysis are inconsistent: For some of the temperature changes, the building-wide power consumption correlates positively with the temperature, while for others, it correlates negatively.

The comparison between Figure 14 and Figure 15 makes this plainly clear: The two temperature changes followed in quick succession on 24 June and 26 June 2025, respectively. Both changes were by one degree centigrade and took place around noon; to the extent that the analysis might be affected by daily cycles (despite the short window), the two should be influenced similarly. Additionally, the local weather on the two days was very similar, sunny and with a 2-degree only maximum temperature difference between the two days.

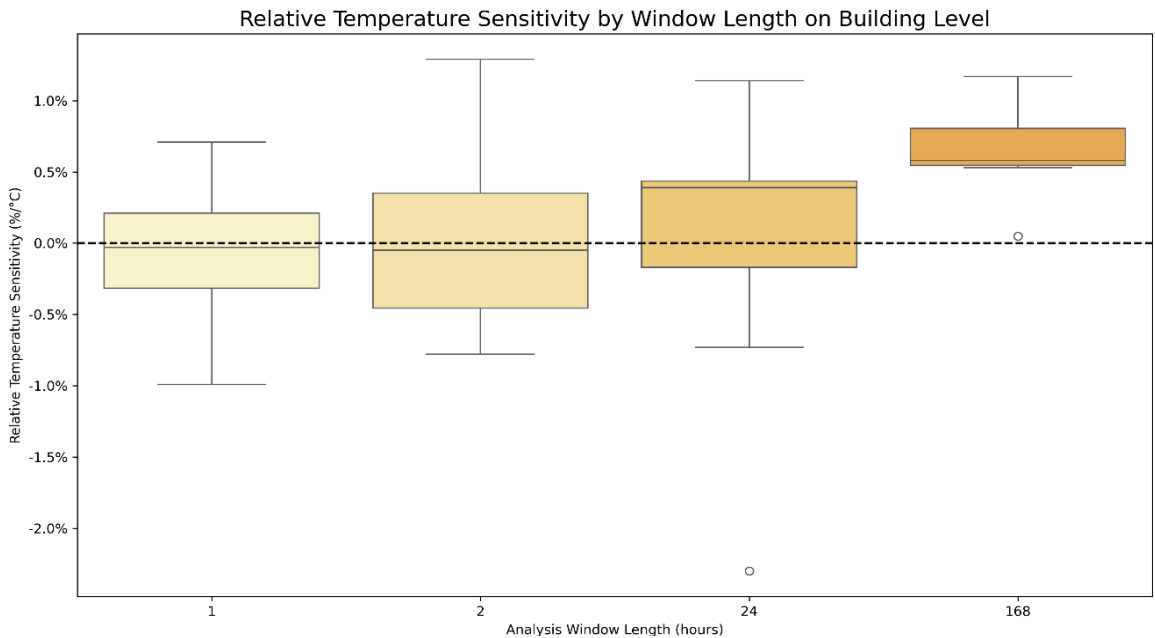

Figure 16: Temperature sensitivity of the building-wide power consumption across the 7 inlet temperature changes, grouped by window size.

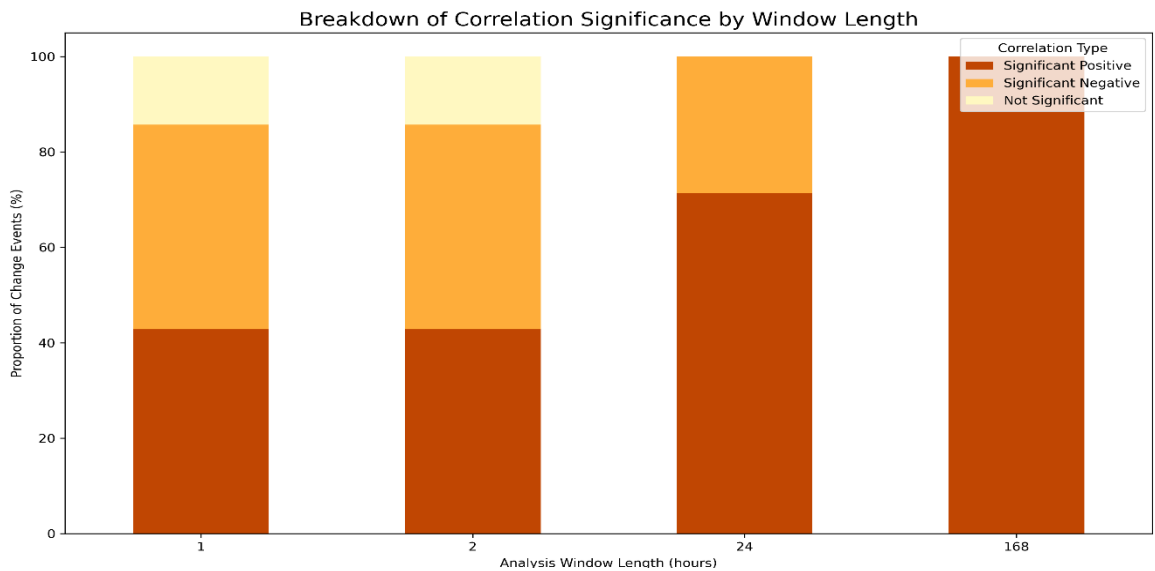

Figure 17: Direction and significance of the correlation between changes in inlet temperatures and changes of the overall building consumption, grouped by size of the analysis window.

Yet, for the change on 24 June there is a significant positive correlation with quite strong positive sensitivity of 1.29 % / °C, while on 26 June, the correlation is negative, equally significant, with quite a substantial (but negative) sensitivity of -0.78 % / °C as well. The results of the 1-hour windows (not shown in the figures) are also both significant and of opposite direction, with respective sensitivities of 0.71 and -0.41 % / °C.

Overall, for each of the (as argued above, likely more relevant) shorter windows of 1 and 2 hours, among the 7 temperature changes, 3 correlations are significant and positive, 3 are significant and negative, while 1 is not significant. Figure 16 shows the distribution of the corresponding temperature sensitivities, while Figure 17 shows the direction and significance of the correlations, each of them grouped by size of the analysis window.

Moreover, the second type of building-level analysis – matching consumption across identical day of week and time of day – revealed similar ambiguity. While some analyses showed overall a positive correlation between temperature and building-wide power consumption, others showed a negative such correlation.

The plots of these analyses look, of course, different: There is no continuous data to follow for the length of a certain analysis window before and after the temperature change. Instead, there are two temporally disjunct, relatively short windows of analysis, which lie a few days before and after the temperature change event, respectively. The distribution of individual consumption values in each of these windows are thus represented in one boxplot each.

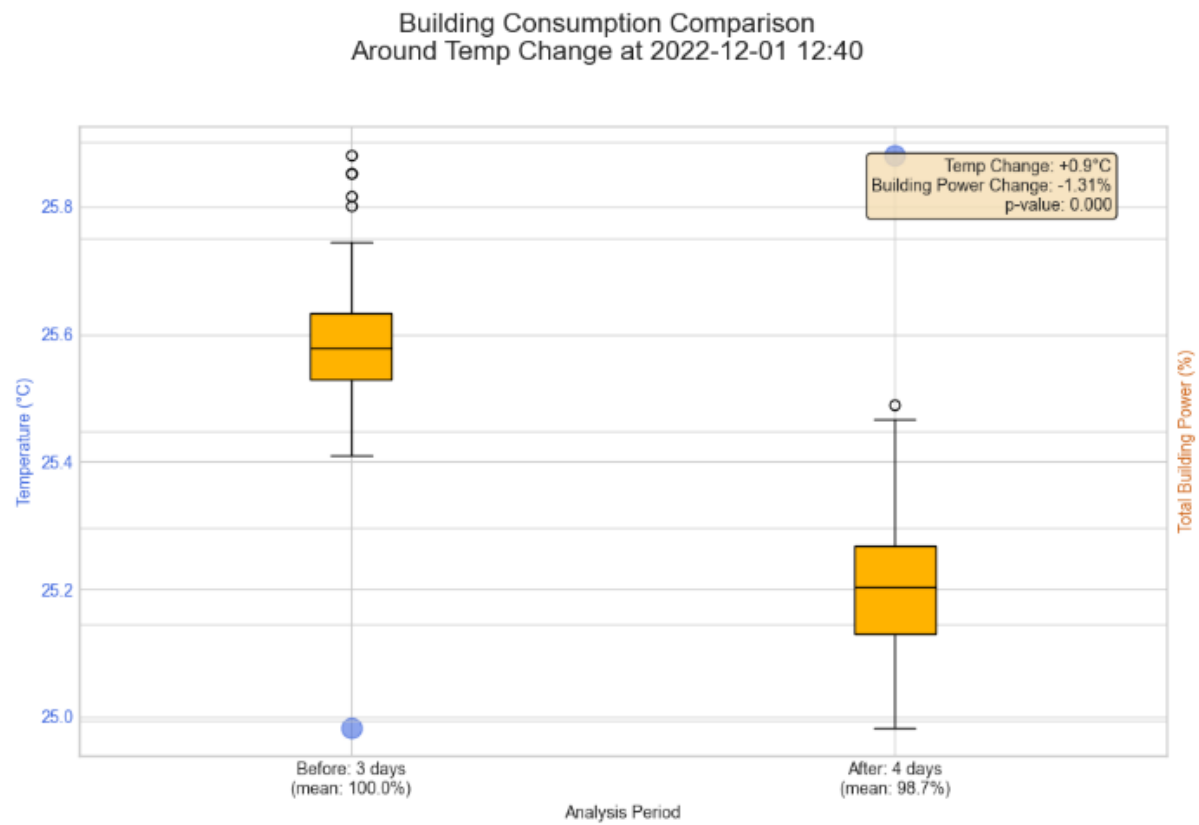

Figure 18: Example for a significant negative correlation between inlet temperatures and overall building power consumption, as analysed in two 1-hour windows exactly 1 week apart, one before and the other after the event.

Figure 18 and Figure 19 underline the ambiguity of these results by showcasing examples of negative and positive correlations between inlet temperatures and overall building power consumption. The December 2022 example in Figure 18 is after a temperature raise of 1 °C in 4 server rooms, covering more than 40% of the DCs power. Although the script always refers to one temperature change event, the June 2025 example actually covers two temperature raises by 1 degree each, on 24 and 26 June, respectively, from 26 to 27 °C first, and then from 27 to 28 °C. They were performed in two server rooms, covering about 20% of the DCs IT power consumption.

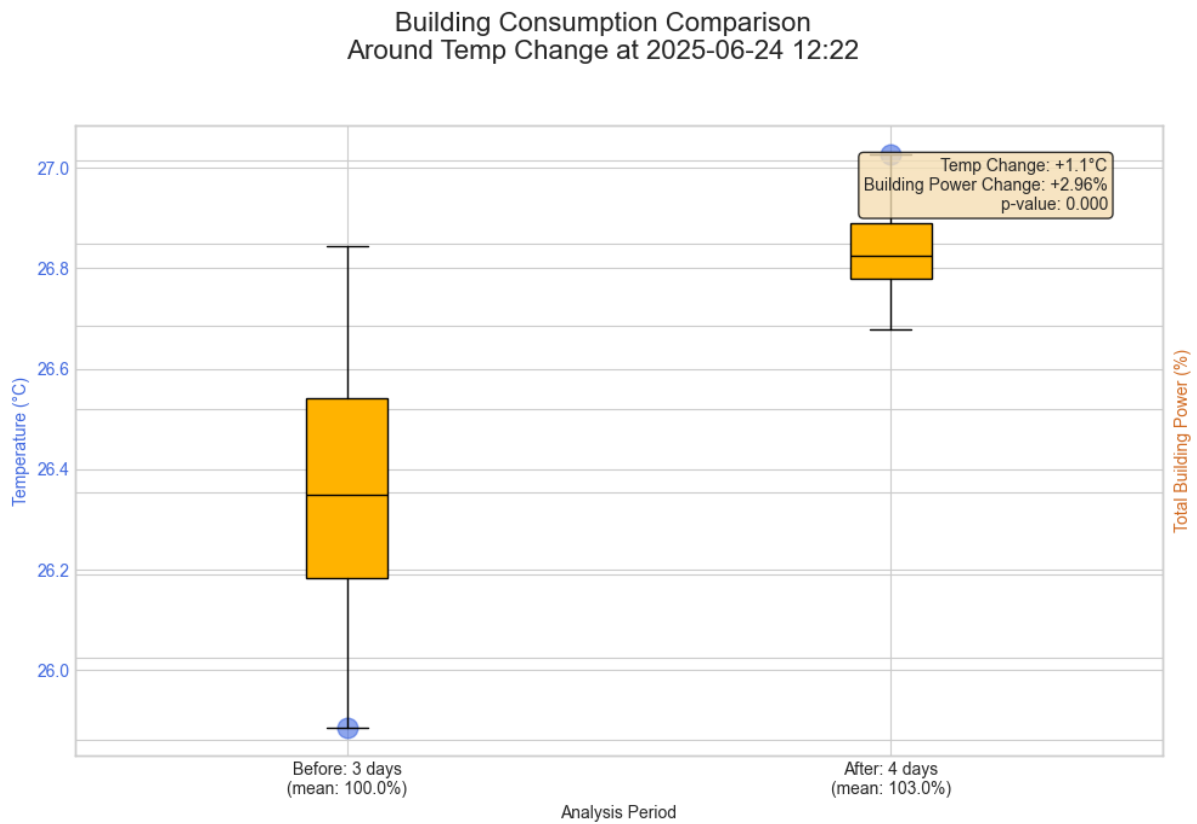

Figure 19: Example for a significant positive correlation between inlet temperatures and overall building power consumption, as analysed in two 1-hour windows exactly 1 week apart, one before and the other after the event.

The effects of the two (1 degree raise for about 40% of the building, 2 degrees raise for around 20% of the building) should have relatively similar effects; yet they yield opposite consequences. This happens throughout the data and sometimes even for the same temperature change event, when comparing different before and after days: Comparing, for example, the power consumptions on the Thursdays before and after the event, might yield the opposite result to comparing the Fridays.

# 7 Conclusions and outlook

This study set out to answer several questions about the relation between server inlet temperatures and the power and energy consumption of data centres, mainly:

- Raising temperatures from today's industry standard of 24 – 25 °C for colocation providers towards the upper limit of the ASHRAE recommendation (i.e., 27 °C) or even above it will indeed lead to an increased consumption of server fans? By how much?

- What will the corresponding cooling infrastructure savings be, and will they compensate for the increased fan consumption, so the net effect will be beneficial? Which are the ideal server inlet temperatures that minimise the overall power consumption?

- Is the PUE a helpful measure for this assessment?

Methodologically, the analysis was based on academic and industry literature research, a few interviews, and the case study of a colocation provider in Switzerland, which provided valuable primary (i.e., measured) data for the study.

A first theoretical background analysis provided the foundations for the subsequent analysis. Among several conceptual insights, the two main learnings from this phase were foundations of data centre cooling (presented in Section 2) and a detailed theoretical understanding of the trade-off between the energy required by the cooling infrastructure and the one of server fans (discussed in Section **Error! Reference source not found.**). Equations 2 – 6 in Section **Error! Reference source not found.** laid out how increasing inlet temperatures require an increased airflow in an air-cooled server room, how this is typically achieved through higher server fan speeds, and how the power consumption of fans is proportional to the cube of their speed.

This theoretical expectation was confirmed through primary data from both DCs of our case study partner, who raised inlet temperatures in two of its data centres from the widespread 24 °C towards the upper limit of the ASHRAE range of 27 °C, and sometimes above. Data from its first data centre was rich, containing inlet temperatures and server room power consumption for 11 server rooms and covering a total of 65 temperature changes. Despite the not ideal resolution, which required harmonisation through downsampling, the analysis could nevertheless establish a significant overall positive correlation between inlet temperature and server power consumption, as discussed in Section 6.1.4.

Data from the second data centre was not so broad, but more focused and with a much better time resolution. It covers only 7 temperature changes in 6 server rooms, but those changes happened simultaneously in 2 – 4 server rooms. Temperature and consumption data were available for each minute, and not for every hour, as the (downsampled) data for the first DC. This analysis confirmed and strengthened the results: For all 7 temperature changes, the most relevant analyses (with time windows of 1 and 2 hours, respectively) all revealed positive and significant correlations between inlet temperature and server power consumption, as shown in Section 6.2.3.

Taken together, the two analyses were also able to quantify the temperature sensitivity of server power consumption: For this case study, and given the more precise data from the second data centre, the likely range of this sensitivity is 0.35 – 0.5 % / °C. In other words, for every degree temperature raise, the server power consumption can be expected to grow by 0.35 – 0.5%. As Figure 7 and the analysis of the second DC have shown, these values were fairly constant in the upper range of the ASHRAE recommendation range and slightly above it, i.e., between 23 – 30 °C.

Part of the data received from the case study partner is sensitive. The absolute power consumptions in particular could not be publicly revealed. Other than for transparency, these numbers, however, would not have been valuable beyond this very case study. The generically important values are the relative changes that were discussed above. The absolute values that are relevant are the temperatures, as they can inform practitioners, researchers, and policymakers alike. These were openly presented throughout the study.

As for the trade-off between server fan and cooling infrastructure power consumptions, the study was unfortunately not able to provide a definitive answer. The literature review presented in Section 5 presents evidence and arguments for both beneficial and detrimental overall effects. And the building-wide primary data analysis for the second DC presented in Section 6.2.4 also yields conflicting evidence.

This ambiguity might rely on the influence of noise, spurious effects such as temporarily turning on or off the chillers and confounding factors, such as variations in the data centre computing load. From our analysis, these effects seem more relevant on the building level than they are for single server rooms, where results were unambiguous.

A possible way forward to address the all-important question of this trade-off could be a controlled experiment. If all (or a substantial share) of a DCs server rooms were to simultaneously go through a couple of inlet temperature changes, while the computing loads were controlled and kept constant throughout the experiment (thus eliminating what this analysis revealed to be the likely most important confounding factor) while at the same time fine-granularly monitoring the power consumption on building level – or better yet, both within server rooms and directly the power consumption of chillers and other cooling components – this could yield significant and possibly unambiguous results.

For the moment, the evidence from hyperscale DCs presented in Section 5.2 shows that DC operators use server inlet temperatures in the range of 27 – 29 °C. As opposed to colocation operators, hyperscalers bear both the costs of server and of infrastructure energy, so have an inherent interest to minimise their sum. This suggests that in their case, the ideal inlet temperatures are at the higher end of the ASHRAE recommendations, or 1-2 degrees above it. At the same time, hyperscale DCs typically deploy server architectures that are highly customised, integrated with, and optimised for the infrastructure of their DCs.

As colocation providers do not have the same level of control over the servers installed in their DCs, it is likely that lower temperatures represent the sweet spot in their case; possibly around the upper end of the ASHRAE recommendation or slightly below it, i.e., 25 – 27 °C. This would also corroborate with the evidence mentioned by Martin Casaulta in the interview.

Meanwhile, it is clear that the PUE is not a helpful metric in this case. This becomes plainly clear when writing PUE as in Equation 7:

$$PUE = 1 + \frac{P_{non-IT}}{P_{IT}}$$

Semantically, the power consumption of server fans should appear in the upper part of the fraction. Because it is difficult to distinguish it from the power consumption of the servers themselves, however, it appears in its lower part. In this context, an inlet temperature increase will thus tend to decrease the upper part of the fraction (because of cooling energy savings) and increase its lower part because of the increased server fan consumption (and instead of increasing the upper part, as would be semantically correct). The overall effect will thus necessarily be a PUE decrease, irrespective of the actual overall consumption impact of the temperature raise. In this context, PUE is thus not only not helpful, but it can easily become misleading.

# References

Alpine Intel (2025) 'The Basics of Chillers', *Alpine Intel*. Available at: https://alpineintel.com/resource/the-basics-of-chillers/ (Accessed: 26 June 2025).

ASHRAE TC9.9 (2016) *Data Center Power Equipment Thermal Guidelines and Best Practices*. American Society of Heating, Refrigerating and Air-Conditioning Engineers, p. 60. Available at: https://www.ashrae.org/File%20Library/Technical%20Resources/Bookstore/ASHRAE_TC0909_Power_White_Paper_22_June_2016_REVISED.pdf.

ASHRAE TC9.9 (2019) *Water-Cooled Servers. Common Designs, Components, and Processes*. American Society of Heating, Refrigerating and Air-Conditioning Engineers, p. 60. Available at: https://www.ashrae.org/File%20Library/Technical%20Resources/Bookstore/WhitePaper_TC099-WaterCooledServers.pdf.

Beaty, D. and Lintner, W. (2008) 'Achieving Energy-Efficient Data Centers with New ASHRAE Thermal Guidelines'. Available at: http://ave.dee.isep.ipp.pt/~mbm/PROJE-EPS/1011/ASHRAE_EEE_2008.pdf.

Cadence System Analysis (2022) 'The Thermal Resistance of Convection'. Available at: https://resources.system-analysis.cadence.com/blog/msa2022-the-thermal-resistance-of-convection (Accessed: 11 May 2025).

Casaulta, M. (2025) 'Server temperatures interview'.

Coroamă, V.C. *et al.* (2025) *Energy Efficiency of Servers: Past and Possible Future Trends*. IEA 4E TCP Efficient, Demand Flexible Networked Appliances (EDNA). Available at: https://www.iea-4e.org/wp-content/uploads/2025/05/EDNA-EE-of-servers-FINAL.pdf.

El-Sayed, N. *et al.* (2012) 'Temperature management in data centers: why some (might) like it hot', in *Proceedings of the 12th ACM SIGMETRICS/PERFORMANCE joint international conference on Measurement and Modeling of Computer Systems*. New York, NY, USA: Association for Computing Machinery (SIGMETRICS '12), pp. 163–174. Available at: https://doi.org/10.1145/2254756.2254778.

EnergieSchweiz (2022) *Mehr ist weniger: Kühlen Sie Ihre Server nicht unter 27 °C*. SFOE. Available at: https://pubdb.bfe.admin.ch/de/publication/download/11119.

European Commission (2019) *Commission Regulation (EU) 2019/424 of 15 March 2019 laying down ecodesign requirements for servers and data storage products pursuant to Directive 2009/125/EC of the European Parliament and of the Council and amending Commission Regulation (EU) No 617/2013 (Text with EEA relevance.)*, *OJ L*. Available at: http://data.europa.eu/eli/reg/2019/424/oj/eng (Accessed: 5 September 2024).

IBM (2015) *Effectively applying the expanded ASHRAE guidelines in your data center*. Available at: https://www.ibm.com/downloads/cas/1Q94RPGE.

Jakob, M., Müller, J. and Altenburger, A. (2021) *Rechenzentren in der Schweiz – Stromverbrauch und Effizienzpotenzial*, p. 100. Available at: https://www.newsd.admin.ch/newsd/message/attachments/66075.pdf.

Judge, P. (2023) 'Over-cooling is past its sell-by date', *Data Center Dynamics*, 31 January. Available at: https://www.datacenterdynamics.com/en/analysis/over-cooling-is-past-its-sell-by-date/ (Accessed: 13 May 2025).

McMordie, R.K. (ed.) (2012) *Solar Energy Fundamentals*. New York: River Publishers. Available at: https://doi.org/10.1201/9780203739204.

Miller, R. (2008) 'Google: Raise Your Data Center Temperature', *Data Center Knowledge*, October. Available at: https://www.datacenterknowledge.com/hyperscalers/google-raise-your-data-center-temperature (Accessed: 14 May 2025).

Moss, D. (2009) *Data Center Operating Temperature: What does DELL Recommend?* Available at: https://i.dell.com/sites/content/business/solutions/whitepapers/en/Documents/dci-Data-Center-Operating-Temperature-Dell-Recommendation.pdf.

Nvidia (2024) *Nvidia DGX B200 Specifications*, *NVIDIA*. Available at: https://www.nvidia.com/en-us/data-center/dgx-b200/ (Accessed: 6 September 2024).

Patel, D. *et al.* (2025) 'Datacenter Anatomy Part 2 – Cooling Systems – SemiAnalysis', *SemiAnalysis*, 13 February. Available at: https://semianalysis.com/2025/02/13/datacenter-anatomy-part-2-cooling-systems/ (Accessed: 10 April 2025).

Patterson, M.K. (2008) 'The effect of data center temperature on energy efficiency', in *Proceedings of the 11th Intersociety Conference on Thermal and Thermomechanical Phenomena in Electronic Systems (ITHERM) 2008. 11th Intersociety Conference on Thermal and Thermomechanical Phenomena in Electronic Systems (ITHERM) 2008*, Orlando, Florida, US, pp. 1167–1174. Available at: https://doi.org/10.1109/ITHERM.2008.4544393.

Petithuguenin, L., Clinger, J. and Leddy, T. (2023) *Study for the review of Commission Regulation 2019/424 (Ecodesign of servers and data storage products)*. ICF, p. 69. Available at: https://eco-servers-review.eu/wp-content/uploads/2023/09/Phase-1-Technical-Analysis-Servers-Review-Study-.pdf.

Rose, B. and Sky, S. (2019) 'Important Parameters for Optimizing DC Fan Operation', *DigiKey*, 30 December. Available at: https://www.digikey.com/en/articles/important-parameters-for-optimizing-dc-fan-operation (Accessed: 11 May 2025).

Sarkinen, J. *et al.* (2020) 'Experimental Analysis of Server Fan Control Strategies for Improved Data Center Air-based Thermal Management', in *2020 19th IEEE Intersociety Conference on Thermal and Thermomechanical Phenomena in Electronic Systems (ITherm)*. *2020 19th IEEE Intersociety Conference on Thermal and Thermomechanical Phenomena in Electronic Systems (ITherm)*, pp. 341–349. Available at: https://doi.org/10.1109/ITherm45881.2020.9190337.

Schelling, P.K., Shi, L. and Goodson, K.E. (2005) 'Managing heat for electronics', *Materials Today*, 8(6), pp. 30–35. Available at: https://doi.org/10.1016/S1369-7021(05)70935-4.

Smith, R. (2024) *NVIDIA Blackwell Architecture and B200/B100 Accelerators Announced: Going Bigger With Smaller Data*. Available at: https://www.anandtech.com/show/21310/nvidia-blackwell-architecture-and-b200b100-accelerators-announced-going-bigger-with-smaller-data (Accessed: 6 September 2024).

Snurr, R. and Freude, D. (2024) 'Physics of Wind Turbines', *Energy Fundamentals*, September. Available at: https://home.uni-leipzig.de/energy/energy-fundamentals/15.htm (Accessed: 11 May 2025).

Swinhoe, D. (2024) 'Hot water, cold water', *Data Center Dynamics*, 31 July. Available at: https://www.datacenterdynamics.com/en/analysis/hot-water-cold-water/ (Accessed: 11 May 2025).

Torrell, W., Brown, K. and Avelar, V. (2016) *The Unexpected Impact of Raising Data Center Temperatures*. 221. Schneider Electric. Available at: https://www.se.com/dk/da/download/document/SPD_VAVR-9SZM5D_EN/ (Accessed: 18 February 2025).

Upsite Technologies (2025) 'How Fan Affinity Laws Impact Fan Energy Savings', *Upsite Technologies - Data Center Cooling Optimization*, 7 May. Available at: https://www.upsite.com/blog/how-fan-affinity-laws-impact-fan-energy-savings/ (Accessed: 11 May 2025).

Vaia (2023) *Thermal Boundary Layer: Equation, Heat Transfer & Thickness*, *Vaia*. Available at: https://www.vaia.com/en-us/explanations/engineering/engineering-thermodynamics/thermal-boundary-layer/ (Accessed: 11 May 2025).